\documentclass[a4paper,12pt]{article}
\pdfoutput=1
\usepackage{geometry}
\usepackage{amsmath,amssymb,amsfonts}
\usepackage{xcolor,graphicx,cite,soul}
\usepackage{array,booktabs,longtable}
\usepackage{caption,microtype}
\usepackage{subcaption}
\usepackage{hyperref}
\usepackage{datetime}
\usepackage{appendix}
\usepackage{ulem}
\usepackage{multirow}  
\usepackage{cancel}
\usepackage{float}



\newcommand{\lsim}
{\;\raisebox{-.3em}{$\stackrel{\displaystyle <}{\sim}$}\;}
\newcommand{\gsim}
{\;\raisebox{-.3em}{$\stackrel{\displaystyle >}{\sim}$}\;}

\newcommand\hotf{\ensuremath{h_{125}}}

\newcommand\ReDiag{\mathop{%
  \raise .5pt\hbox{[}%
  \widetilde{\mathrm{Re}}%
  \raise .5pt\hbox{]}}}
\newcommand\ReOffDiag{\mathop{%
  \raise .5pt\hbox{$\llbracket$}%
  \widetilde{\mathrm{Re}}%
  \raise .5pt\hbox{$\rrbracket$}}}

\newcommand\Mh{m_h}
\newcommand\MH{m_H}

\newcommand\refeq[1]{Eq.~(\ref{#1})}
\newcommand\refeqs[1]{Eqs.~(\ref{#1})}
\newcommand\refta[1]{Tab.~\ref{#1}}
\newcommand\refse[1]{Sect.~\ref{#1}}

\newcommand\citere[1]{Ref.~\cite{#1}}
\newcommand\citeres[1]{Refs.~\cite{#1}}

\newcommand\wrt{w.r.t.\ }

\newcommand{\CP}{{\cal CP}}
\newcommand{\cp}{{\CP}}

\newcommand{\tev}{\,\, \mathrm{TeV}}
\newcommand{\gev}{\,\, \mathrm{GeV}}

\newcommand{\eeZhh}{\ensuremath{e^+e^- \to Z h h}}

\newcommand\HB{\texttt{HiggsBounds}}

\newcommand\fb{\ensuremath{\,\mbox{fb}}}
\newcommand\ab{\ensuremath{\,\mbox{ab}}}

\newcommand\iab{\ensuremath{\ab^{-1}}}

\newcommand{\br}{\text{BR}}

\newcommand{\sig}{\sigma}
\newcommand{\sigSM}{\sig_{hh}^{\rm SM}}
\newcommand{\sigRx}{\sig_{hh}^{\rm RxSM}}
\newcommand{\sigeeSM}{\ensuremath{\sig_{\rm SM}}}
\newcommand{\sigeeRx}{\ensuremath{\sig_{\rm RxSM}}}
\newcommand{\sigeeNoH}{\ensuremath{\sig_{\rm NoH}}}
\newcommand{\sigeeH}{\ensuremath{\sig_{\rm H}}}

\def\reffi#1{\mbox{Fig.~\ref{#1}}}
\def\reffis#1{\mbox{Figs.~\ref{#1}}}

\def\Ga{\Gamma}

\def\ga{\gamma}

\def\la{\lambda}
\newcommand\kala{\ensuremath{\kappa_{\lambda}}}
\newcommand\laSM{\ensuremath{\lambda_{\mathrm{SM}}}}
\newcommand{\lahhh}{\ensuremath{\la_{hhh}}}
\newcommand{\lahhH}{\ensuremath{\la_{hhH}}}
\newcommand{\lahHH}{\ensuremath{\la_{hHH}}}

\newcommand{\laHHH}{\ensuremath{\la_{HHH}}}

\newcommand{\mhh}{\ensuremath{m_{hh}}}

\definecolor{Orange}{named}{orange}
\definecolor{Purple}{named}{purple}
\definecolor{Lightblue}{cmyk}{0.9,0.1,0.1,0.3}
\definecolor{dgelborange}{cmyk}{0.,0.3,0.5, 0.}
\definecolor{Lila}{rgb}{0.5,0.,1}
\definecolor{Darkgreen}{rgb}{0.,.7,0.2}

\newcommand{\vS}{x}

\graphicspath{{figs/}}
\allowdisplaybreaks
\begin{document}
\thispagestyle{empty}

\def\thefootnote{\fnsymbol{footnote}}

\begin{flushright}
\mbox{}
DESY-24-213\\
IFT--UAM/CSIC-24-186\\
KA-TP-01-2025 \\
\end{flushright}


\begin{center}

{\large\sc 
{\bf Sensitivity to Triple Higgs Couplings via Di-Higgs Production\\[.5em]
  in the RxSM at the (HL-)LHC and future \boldmath{$e^+e^-$} Colliders}}  

\vspace{1cm}

{\sc
F.~Arco$^{1}$%
\footnote{emails:
francisco.arco@desy.de,
Sven.Heinemeyer@cern.ch,
margarete.muehlleitner@kit.edu,\\
\mbox{}\hspace{19mm}andrea.parra@estudiante.uam.es, nestoriv@ucm.es,
Alain.Verduras@desy.de} 
, S.~Heinemeyer$^{2}$%
, M.~M\"uhlleitner$^{3}$%
,\\[.5em] A.~Parra Arnay$^{2}$%
, N.~Rivero Gonz\'alez$^{4}$%
~and A.~Verduras Schaeidt$^{1}$%
}

\vspace*{.7cm}

{\sl
$^1$Deutsches Elektronen-Synchrotron DESY, Notkestr.\ 85, 22607 Hamburg,
Germany  

\vspace{0.1em}

$^2$Instituto de F\'isica Te\'orica (UAM/CSIC), 
Universidad Aut\'onoma de Madrid, \\ 
Cantoblanco, 28049, Madrid, Spain

\vspace*{0.1cm}

$^3$Institute for Theoretical Physics,
Karlsruhe Institute of Technology, 76128, Karlsruhe, Germany

\vspace*{0.1cm}

$^4$ Departamento de F\'isica Te\'orica, Universidad Complutense de Madrid, 28040 Madrid, Spain

}

\end{center}

\vspace*{0.1cm}

\begin{abstract}
\noindent
The real Higgs singlet extension of the Standard Model (SM) without $Z_2$ symmetry, the RxSM, 
is the simplest extension of the SM that features a First Order Electroweak Phase Transition 
(FOEWPT) in the early universe. The FOEWPT is one of the requirements needed for electroweak 
baryogenesis  to explain the baryon asymmetry of the universe (BAU). Thus, the RxSM is a perfect 
example to study features related to the FOEWPT at current and future collider experiments. 
The RxSM has two $\cp$-even Higgs bosons, $h$ and $H$, with masses $\Mh < \MH$, where we assume
that $h$ corresponds to the Higgs boson discovered at the LHC. Our analysis is based on a benchmark 
plane that ensures the occurence of a strong FOEWPT, where $\MH > 2 \Mh$ is found. In a first 
step we analyze the di-Higgs production at the (HL-)LHC, $gg \to hh$, with a focus on the impact
of the trilinear Higgs couplings (THCs), \lahhh\ and \lahhH. The interferences of the 
resonant $H$-exchange diagram involving \lahhH\ and the non-resonant diagrams result in a characteristic
peak-dip (or dip-peak) structure in the \mhh\ distribution. We analyze how \lahhH\ can be accessed, 
taking into account the experimental smearing and binning. 
We also demonstrate that the approximation used by ATLAS and CMS for the resonant di-Higgs searches
may fail to capture the relevant effects and lead to erroneous results.
In a second step we analyze the benchmark plane at a future high-energy $e^+e^-$ collider with 
$\sqrt{s} = 1000 \gev$ (ILC1000). 
We demonstrate the potential sensitivity to \lahhH\ via an experimental 
determination at the ILC1000.
\end{abstract}


\def\thefootnote{\arabic{footnote}}
\setcounter{page}{0}
\setcounter{footnote}{0}

\newpage


\section{Introduction}
\label{sec:intro}

\noindent
The discovery of a new scalar particle with a mass of $\sim125\gev$
by ATLAS and CMS~\cite{Aad:2012tfa,Chatrchyan:2012xdj,Khachatryan:2016vau}
is in agreement
--- within the experimental and theoretical uncertainties ---
with the predictions for the properties of the Standard Model (SM) Higgs boson.
Furthermore, no conclusive sign of Higgs bosons beyond the SM
(BSM) has been observed so far.
The experimental results for the couplings of the Higgs boson discovered at the LHC
with a mass of about $\sim 125 \gev$, \hotf, 
are known up to now to an experimental precision of  
roughly $\sim 10-20\%$, leave ample room for interpretations in BSM
models.  
On the other hand, the SM fails to address various major existing observations in nature. 
One of the most interesting open questions
concerns the origin of the matter-antimatter asymmetry of the Universe, which
(according to the measured value of the mass of the Higgs boson) cannot be explained 
in the SM~\cite{Kajantie:1996mn}.
Models featuring extended
Higgs sectors could allow for the generation of the baryon asymmetry of the Universe (BAU) via
electroweak (EW) baryogenesis~\cite{Kuzmin:1985mm}.
Here, a key ingredient is the realization of a First Order Electroweak Phase Transition (FOEWPT) that
can take place in models with BSM Higgs sectors in the early universe when a tunneling
from one minimum of the Higgs potential to a deeper one occurs~\cite{bsmpt1,bsmpt2,bsmpt3,bsmpt4,bsmpt5,bsmpt6,bsmpt7,bsmpt8}. 
Such a FOEWPT fulfills one of the three Sakharov conditions required for EW baryogenesis~\cite{sakharov}. 
Consequently, one of the main tasks of present and future
colliders will be to determine whether  
the observed scalar bosons part of the Higgs sector of an extended
model. 

In contrast to the Higgs couplings to the SM third generation
fermions and to the gauge bosons, the trilinear  
Higgs self-coupling $\lahhh$ remains to be determined, where
we will use the abbreviation $\kala \equiv \lahhh/\laSM$ in the following to quantify possible deviations from the SM value $\laSM$. 
So far it has been constrained by ATLAS~\cite{ATLAStrilinear} to be inside the
range $-1.2 < \kala < 7.2$ at the 95\% C.L.\
and $-1.39 < \kala < 7.02$ at the 95\% C.L.~by
CMS~\cite{CMStrilinear}, 
both assuming a SM-like top-Yukawa coupling of the discovered Higgs boson at 125~GeV.
Many BSM models can still induce significant deviations in the
trilinear coupling $\lahhh$ of the $125 \gev$ Higgs boson with respect
to the SM value, while all other experimental and theoretical constraints 
are fulfilled, see, e.g., \citere{Abouabid:2021yvw} for a recent overview. 
For reviews on the measurement of the triple Higgs couplings
at future colliders see for instance \citeres{deBlas:2019rxi, DiMicco:2019ngk}.
In case a BSM Higgs sector manifests itself, it will be
a prime task to measure not only $\lahhh$, but also the other BSM triple Higgs
couplings (THCs) that can be realized in the model. First studies for the (HL-)LHC in this direction 
can be found in \citeres{Arco:2022lai,Heinemeyer:2024hxa}.

One of the simplest extensions of the SM Higgs sector is the 
Higgs-Singlet extension of the SM (RxSM)~\cite{model,model2,Costa:2015llh,paper}.
A real singlet, ${\cal S}$, is added to the SM Higgs sector with the doublet ${\cal H}$,  
leading to two physical Higgs bosons after electroweak symmetry breaking. We will assume that the lighter one, $h$, 
corresponds to the Higgs-boson observed at the LHC, whereas the heavier one, $H$, 
has escaped detection so far. 
The model can be realized with or without a $Z_2$ symmetry, under which 
${\cal H} \to {\cal H}$ and ${\cal S} \to -{\cal S}$. If no $Z_2$ symmetry is imposed,
a linear and a cubic term in ${\cal S}$ is allowed. In \citere{bsmpt1} 
it was shown that in this case the RxSM can exhibit a FOEWPT in the early universe. 
Consequently, the RxSM constitutes the simplest Higgs-sector extension featuring
a FOEWPT. In \citeres{paper,bsmpt2,bsmpt4} 
it was demonstrated that within the RxSM this can only be realized if the mass of the second Higgs boson is not too large,
$\MH \lsim 900 \gev$. For a broad class of models this was shown in \citeres{paper,LHCGW}.
This makes the boson $H$ a prime target for the HL-LHC, but also for
future high-energy $e^+e^-$ colliders. 

The related physics questions are two-fold. Can such a new, heavier Higgs boson, 
as favored by a FOEWPT, be detected at the HL-LHC and/or high-energy $e^+e^-$ colliders?
Can we gain access to BSM THCs through the processes 
$gg \to hh$ at the HL-LHC, or $e^+e^- \to Zhh/\nu\bar \nu hh$ at a high-energy $e^+e^-$ collider, via a resonant
$s$-channel heavy Higgs boson exchange, involving the BSM THC $\lahhH$? 

In our analysis we will assume the RxSM exhibiting a FOEWPT in the early universe
as required by EW baryogenesis. Here we make use of \citere{paper} defining a 2-dimensional
benchmark plane that allows an analysis where the FOEWPT is ensured.  
We will furthermore assume that the additional Higgs boson 
$H$ will have been found at the HL-LHC in single heavy Higgs production. 
For the HL-LHC we will analyze di-Higgs production, $gg \to hh$, as evaluated with the 
code {\tt HPAIR}~\citeres{HP2,Dawson:1998py}, 
where the $H$ appears as a heavy $s$-channel resonance, 
$gg \to H \to hh$. Employing the invariant di-Higgs mass distributions, $\mhh$, we will 
analyze the sensitivity of the HL-LHC to the BSM THC $\lahhH$. In passing, we will 
demonstrate that the simplification employed by ATLAS and CMS in the search for
resonant di-Higgs production, leaving out the SM-like contributions in the signal model,
is highly questionable (see also \citere{Heinemeyer:2024hxa}.) 

The analysis is 
extended to future high-energy $e^+e^-$ colliders, like the ILC~\cite{Bambade:2019fyw}
and CLIC~\cite{Charles:2018vfv}, which can play a key role for the
measurement of the Higgs potential with high precision and in detecting possible deviations
from the SM~\cite{Djouadi:1999gv,Abramowicz:2016zbo,Strube:2016eje,Roloff:2019crr,deBlas:2019rxi,DiMicco:2019ngk}.%
\footnote{
Circular colliders such as FCC-ee or CEPC have only very limited
sensitivity to the SM-like THC via loop effects in single Higgs
production. Furthermore, they do not have sufficient center-of-mass energy
to produce a heavy RxSM Higgs-boson resonantly to gain access to $\lahhH$. 
}%
~More specifically, we will analyze the process $e^+e^- \to ZH \to Z \, hh$ at the ILC1000,
i.e.\ a linear collider operating at $\sqrt{s} = 1000 \gev$.%
\footnote{While we use certain characteristics of the ILC (e.g.\ foreseen luminosities), our results
are valid for any $e^+e^-$ collider operating at the same center-of-mass energy.}%
~Like for the process $gg \to hh$ the heavy Higgs
can appear as a resonance, involving $\lahhH$. Employing the $\mhh$ distributions
we derive the sensitivity to $\lahhH$ at the ILC1000. 
Similar analyses within the 2HDM, but without the focus on the FOEWPT, can 
be found in \citere{Arco:2022lai} for the HL-LHC and in \citeres{Arco:2021bvf,Arco:2022xum} for ILC and CLIC. 
Further analyses involving BSM triple Higgs couplings can be found in
\citeres{Djouadi:1999rca,Basler:2017uxn,Basler:2019iuu,DiMicco:2019ngk,Abouabid:2021yvw}. 

The RxSM has been analyzed during the last years w.r.t.\ FOEWPTs and related phenomena.
As discussed above, in \citere{paper} the occurance of a FOEWPT in the RxSM connected to di-Higgs production was analyzed,
followed up by a more detailed analysis of the process $hh \to b \bar b \ga\ga$ in \citere{Zhang:2023jvh}. 
A similar analysis focusing on the $b\bar b ZZ$ final state was presented in \citere{Palit:2023dvs}. 
The interplay of EW bayrogenesis with gravitational waves in the RxSM was first analyzed in \citere{Ellis:2022lft},
see also \citeres{Niemi:2024vzw,Gould:2024jjt} for the inclusion of two-loop effects into the evaluation of the FOEWPT.
The inclusion of dimension-6 operators on top of a one-loop calculation was analyzed in \citere{Giovanakis:2024rvg}.
Results for the FOEWPT employing lattice calculations were presented in \citeres{Niemi:2024axp,Ramsey-Musolf:2024ykk}.
The latter paper
also discussed the HL-LHC phenomenology of resonant di-Higgs searches, but taking into account only the resonant 
$H$-exchange contribution. The interference effects with the non-resonant diagrams in the RxSM were analyzed in \citere{Feuerstake:2024uxs},
but without taking into account the possibility of a FOEWPT.
Concerning future colliders, the resonant di-Higgs production in the RxSM at $e^+e^-$ colliders was discussed in \citere{Lewis:2024yvj},
but again without taking the FOEWPT into account. A corresponding analysis for a future $\mu^+\mu^-$ collider can be found in 
\citere{Aboudonia:2024frg}. In contrast to the existing analyses, 
in this work we combine the requirement of a FOEWPT with the analysis of di-Higgs production at the 
HL-LHC including all contributing diagrams, as well as the complementary di-Higgs production at future $e^+e^-$ colliders.

Our paper is organized as follows. In \refse{sec:model} we briefly review
the RxSM, fix our notation, and define the benchmark plane 
featuring a FOEWPT, used
later for our investigation, and summarize the constraints that we
apply. In \refse{sec:hllhc} we analyze the possible sensitivity of the
di-Higgs production cross section at the (HL-)LHC to $\lahhH$
and discuss the applicability of resonant di-Higgs searches by ATLAS
and CMS to specific models such as the RxSM.
In \refse{sec:ilc} we extend our analysis of the $\lahhH$ to the ILC1000. 
Our conclusions are given in \refse{sec:conclusions}. 


\section{The Model and the Constraints}
\label{sec:model}

In this section we give a short description of the RxSM to fix our
notation. We briefly review the theoretical and experimental constraints.
Finally we will define the benchmark plane and the benchmark points
for our analysis of the the di-Higgs production at the (HL-)LHC and the ILC.


\subsection{The RxSM}
\label{sec:rxsm}

\noindent

The model that is the framework of our analysis, is the real singlet extension of the SM (RxSM),
without imposing a $Z_2$ symmetry (see \citeres{paper,constrains,model2} for reviews). This Higgs sector extension
adds a real singlet field ${\cal S}$ to the complex Higgs doublet ${\cal H}$ of the SM Higgs Sector. 
After EWSB the doublet and the singlet fields can be written in the unitary gauge as
\begin{equation}
{\cal{H}}=\left(
 \begin{matrix}
 0  \\
 \frac{h'+v}{\sqrt{2}}
 \end{matrix}\right)\,, \qquad 
 {\cal S}=s+\vS\,, 
 \label{singlet}
\end{equation}
where $h'$ and $s$ are two CP even Higgs fields,
$\vS$ is the singlet vev 
and $v\sim246\gev$ is the SM vev.
The potential is given by, 
\begin{align}
    V({\cal{H}},{\cal S})=-\mu^2({\cal{H}}^{\dagger }{\cal{H}})+\lambda({\cal{H}}^{\dagger }{\cal{H}})^2+\frac{a_1}{2}({\cal{H}}^{\dagger }{\cal{H}}){\cal S}+\frac{a_2}{2}(H^{\dagger }{\cal{H}}){\cal S}^2+\frac{b_2}{2}{\cal S}^2+\frac{b_3}{3}{\cal S}^3+\frac{b_4}{4}{\cal S}^4. \label{potential}
\end{align}
The new potential has seven Lagrangian parameters, 
where $a_1$ and $b_3$ are allowed since no $Z_2$ symmetry is imposed.
Taking into account the vevs of the doublet and the singlet fields the model has nine free parameters. 
The number of free parameters can be reduced to seven by using the minimization conditions of the potential,
\begin{equation}
    \left|\frac{dV}{dh'}\right|_{h'=0,s=0}=0\,,\qquad \left|\frac{dV}{ds}\right|_{h'=0,s=0}=0,
\end{equation}
which leads to two relations between the Lagrangian parameters,
\begin{align}
    \mu^2 &= \lambda v^2+(a_1+a_2\vS)\frac{\vS}{2}, \nonumber\\
    b_2 &= -b_3\vS-b_4\vS^2-\frac{a_1v^2}{4\vS}-\frac{a_2v^2}{2} 
\label{mincon}.
\end{align}
Expanding the potential using the definitions in \refeq{singlet}, the mass matrix can be computed\,as,
\begin{equation}
    M^2= 
    \left.\begin{pmatrix}
\frac{d^2V}{dh'^2} & \frac{d^2V}{dh'ds} \\
 \frac{d^2V}{dh'ds}  & \frac{d^2V}{ds^2}
\end{pmatrix}\right|_{h'=0,s=0} \equiv
\begin{pmatrix}
m_{h'}^2 & m_{h's}^2 \\
 m_{h's}^2  & m_s^2
\end{pmatrix}=
\begin{pmatrix}
2\lambda v^2 & (a_1+2a_2\vS)\frac{v}{2} \\
  (a_1+2a_2\vS)\frac{v}{2}  & b_3\vS+2b_4\vS^2-\frac{a_1v^2}{4\vS}
\end{pmatrix}.\label{massmatrix}
\end{equation}
To obtain the expressions for the masses of the physical states $h$ and $H$, the mass matrix has to be diagonalized with the angle $\theta$,
\begin{equation}
    \begin{pmatrix}
 h \\
  H
\end{pmatrix}=
\begin{pmatrix}
\cos\theta & \sin\theta \\
 -\sin\theta & \cos\theta
\end{pmatrix}
  \begin{pmatrix}
 h' \\
  s
\end{pmatrix}\,.
\label{mixing matrix}
\end{equation}
We define $h$ as the lighter state, which we identify with the Higgs boson discovered at the LHC with a mass of $\sim 125 \gev$.
Correspondingly, $H$ is defined as the second, heavier Higgs boson. The physical masses are given by, 
\begin{equation}
    m_{h,H}^2=\frac{1}{2}\left(m_{h'}^2+m_s^2\mp\left|m_{h'}^2-m_s^2\right|\sqrt{1+\left(\frac{2m_{h's}^2}{m_{h'}^2-m_s^2}\right)^2}\right),\label{massterms}
\end{equation}
and the mixing angle by
\begin{equation}
    \sin{2\theta}=\frac{2m_{h's}^2}{m_H^2-m_h^2}.
\label{mixingangle}
\end{equation}
The various triple Higgs couplings between the physical fields are obtained from the Higgs potential after rotation to 
the mass basis and taking the third derivatives with respect to the physical Higgs fields.
The triple Higgs couplings $\lambda_{h_i h_j h_k}$ between the physical fields $h_i$, $h_j$, and 
$h_k$ ($h_{i,j,k} \in \{h,H\}$) are defined via the Feynman rule 
\begin{equation}
        \begin{gathered}
                \includegraphics{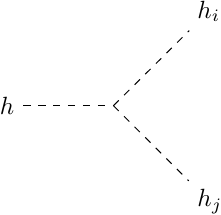}
        \end{gathered}
        =- i\, v\, n!\; \la_{h h_i h_j}\,,
\label{eq:lambda}
\end{equation}
where $n$ is the number of identical particles in the vertex. 
The following expressions are obtained for the couplings $\lahhh$ between the three SM-like fields $h$
(called SM-like coupling in the following) and $\lahhH$ (called BSM coupling in the following),
\begin{align}
    \lahhh &= \frac{1}{v}\left[\left(\frac{a_1}{4}+\frac{a_2\vS}{2}\right)\cos^2\theta\sin\theta+\frac{a_2v}{2}\cos\theta\sin^2\theta+\left(\frac{b_3}{3}+b_4\vS\right)\sin^3\theta+\lambda v\cos^3\theta\right]\,, \\
    \lahhH &= \frac{1}{4v}\Big[(a_1+2a_2\vS)\cos^3\theta+4v(a_2-3\lambda)\cos^2\theta\sin\theta \nonumber\\
    &\quad -2(a_1+2a_2\vS-2b_3-6b_4\vS)\cos\theta\sin^2\theta - 2a_2v\sin^3\theta\Big]. 
    \label{couplings}
\end{align}
There are two more BSM triple Higgs couplings, involving two or three heavy Higgs bosons, $\lahHH$ and $\laHHH$. 
However, they do not play a role in our analysis. The couplings of the Higgs bosons $h$ and $H$, respectively, to any SM particle $i$ 
are modified w.r.t.\ the corresponding SM Higgs-boson coupling $g_{h'i}$. In the RxSM, the couplings are suppressed compared to $g_{h'i}$
by the cosine (sine) of the mixing angle for the light (heavy) Higgs-boson as,  
\begin{equation}
    g^{\rm RxSM}_{hi}=g^{\rm SM}_{h'i}\cdot\cos\theta\,, \qquad 
    g^{\rm RxSM}_{Hi}=g^{\rm SM}_{h'i}\cdot\sin\theta\,. 
    \label{smcouplings}
\end{equation}
Correspondingly, the decay rates of the RxSM Higgs bosons $h$ and $H$, respectively, to the SM particles change as
\begin{equation}
    \Ga^{\rm RxSM}_{h\to ii}=\Gamma_{H_{\rm SM} \to ii}\cdot\cos^2\theta \,, \qquad
    \Ga^{\rm RxSM}_{H\to ii}=\Gamma_{H_{\rm SM} \to ii}\cdot\sin^2\theta \,
\end{equation}
(with $H_{\rm SM}$ denoting the Higgs boson in the SM, and its mass $m_{H_{\rm SM}}$ correspondingly set to 
$\Mh$ or $\MH$ in $\Gamma_{H_{\rm SM} \to ii}$). The total decay width of the heavy Higgs boson
can additionally be modified by the possible decay into two light Higgs-bosons, which (for $\MH > 2 \Mh$) is given by decay width
\begin{equation}
    \Gamma_{H\to hh}=\frac{\lahhH^2}{8\,\pi\,\MH} \, \sqrt{1-\frac{4 \Mh^2}{\MH^2}}\,.
\end{equation}
The total decay width of the heavy Higgs boson is then calculated as
\begin{equation}
    \Ga_H = \Ga_{H_{\rm SM}} \cdot\sin^2\theta + \Ga_{H\to hh}\,.
\end{equation}
For $\cos\theta \to 1$, the light Higgs boson $h$ has the same properties as predicted by the SM with 
$\Ga_{h}=\Gamma_{H_{\rm SM}}$ (for $\Mh = m_{H_{\rm SM}}$), which is known as the alignment limit. 
In case of $\cos\theta \to 0$, $H$ has SM-like couplings and $\Ga_{H} = \Gamma_{H_{\rm SM}}$ 
for $\Gamma_{H\to hh}=0$ (and $\MH = m_{H_{\rm SM}}$). 

Instead of using the original Lagrangian parameters it convenient to employ a more physical parametrization of the RxSM.
In this work we use the ``mass basis'', which includes the masses of the Higgs bosons. It is defined as
\begin{equation}
    \Mh, \, \MH, \, v, \, \vS, \, \theta, \, a_2, \, b_4.
\end{equation}
The transition from the original basis is given by 
\begin{align}
\la &= \frac{1}{2v^2}(\MH^2\sin^2\theta + \Mh^2\cos^2\theta)\,,\label{trans1}\\
b_2 &= \Mh^2\sin^2\theta + \MH^2\cos^2\theta-\frac{a_2}{2}v^2\,,\\
a_1 &= \frac{2}{v}\sin\theta\cos\theta(\MH^2 - \Mh^2)\,.
\label{trans3}
\end{align}
Now we can rewrite the triple Higgs couplings as
\begin{align}
    \lambda_{hhh}=&\frac{1}{6v^2}\{3m_h^2\cos^5\theta + 3 a_2v\vS\cos^2\theta\sin\theta+3\cos\theta[a_2v^2-(m_h^2-2m_H^2)\cos^2\theta]\sin^2\theta\nonumber\\
    &+2v(b_3+3 b_4 \vS)\sin^3\theta\} ,\nonumber\\
    \lambda_{hhH}=&\frac{1}{2v^2}\{a_2v\vS\cos^3\theta+(m_H^2 - 4m_h^2)\sin\theta\cos^4\theta+\cos^2\theta[2a_2v^2\sin\theta\nonumber\\
    &+(2m_h^2-5m_H^2)\sin^3\theta]+v\sin\theta[(b_3-a_2\vS+3b_4\vS)\sin2\theta-a_2v\sin^2\theta]\}  .\label{trilinears}
\end{align}


\subsection{Theoretical and Experimental Constraints}
\label{sec:constraints}

In this subsection we briefly summarize the various theoretical and
experimental constraints considered in our analysis. It should be noted that we did not
check for constraints arising from di-Higgs measurements at the LHC (this will be commented on below, where relevant).

\begin{itemize}

\item {\bf Theoretical constraints}\\
It has to be ensured (we follow \citeres{paper,constrains}) that the potential is bounded from below and that the couplings
are in the perturbative regime. Firstly, for the scalar potential to be stable, the quartic couplings must be positive
in all directions of the fields. To ensure this, we demand that the determinant of the Hessian matrix is positive. 
Second, to ensure perturbative couplings, we require that $\lambda$, $\frac{a_2}{2}$ and $\frac{b_4}{4}$ are smaller than $4\pi$. 
Taking all this into account the following theoretical constraints are found~\cite{paper}:
\begin{equation}
    0<\lambda,\;\frac{a_2}{2}\;\;\&\;\;\lambda,\;\frac{a_2}{2},\;\frac{b_4}{4}<4\pi\;\;\&\;\; a_2>-2\sqrt{\lambda b_4}.
    \label{eq:pert}
\end{equation}
As will be seen later, phenomenological restrictions even yield $b_4 < 1$.
As pointed out in \citere{paper}, unitarity constraints do not yield further restrictions on the parameter space beyond \refeq{eq:pert}.
\item {\bf Constraints from direct Higgs-boson searches at colliders}\\
The exclusion limits at the $95\%$ C.L.\
of all relevant BSM Higgs boson searches (including Run~2 data from the
LHC) are included in the public code
\HB\,\texttt{v.6}~\cite{Bechtle:2008jh,Bechtle:2011sb,Bechtle:2013wla,Bechtle:2015pma,Bechtle:2020pkv,Bahl:2022igd}, which is included in the public code \texttt{HiggsTools}~\cite{Bahl:2022igd}.
~For a parameter point in a particular model, \HB\ determines on the
basis of expected limits which is the most
sensitive channel to test each BSM Higgs boson.
Then, based on this most sensitive channel, \HB\ determines whether the
point is allowed or not at the $95\%$~CL. 
As input \HB\ requires some specific predictions from the model,
like branching ratios or Higgs-boson couplings (which we evaluate as described in \refse{sec:rxsm}).

\item {\bf Constraints from the properties of the \boldmath{$\sim 125 \gev$} Higgs boson}\\
Any model beyond the SM has to accommodate a Higgs boson 
with mass and signal strengths as they were measured at the LHC.
For the parameter points used, the compatibility of the $\cp$-even scalar
$h$ with a mass of $125.09\gev$, $h_{125}$, with the measurements of signal
strengths at the LHC is tested with the code
\texttt{HiggsSignals}\,\texttt{v.3}~\cite{Bechtle:2013xfa,Bechtle:2014ewa,Bechtle:2020uwn,Bahl:2022igd}, 
which is included in the code \texttt{HiggsTools}. 
The code provides a statistical $\chi^2$ for the $h_{125}$
predictions of a given model in comparison to the measurements of the
Higgs-boson signal rates and masses from the LHC.
Specifically, we demand that $\chi^2_{h_{125}{\rm , RxSM}} - \chi^2_{h_{125}{\rm , SM}} < 6.3$ (with $\chi^2_{h_{125}{\rm , SM}} = 159.7$).

\end{itemize}


\subsection{Benchmark Plane and Points}
\label{sec:planes}

After applying the minimization conditions the model has seven free parameters: $a_1,\,a_2,\,b_3,\,b_4,\,\lambda,\,v$ and $\vS$. 
Fixing the SM-like Higgs mass $m_h$ and the SM VEV $v$ to their phenomenological values $\Mh \approx 125 \gev$ and $v \approx 246 \gev$, we are left with five free parameters, $a_1,\,b_3,\,b_4,\, \lambda,$ and $\vS$.
The aim in this subsection is to define a benchmark plane with only two degrees of freedom which features a SFOEWPT,
and which maximizes the di-Higgs production cross section at the LHC. The authors of \citere{paper} scanned the RxSM 
parameter space and kept points exhibiting a SFOEWPT. Out of these they selected eleven points that maximize the di-Higgs
production cross section at the LHC, $\sig(pp \to H \to hh)$. 
We have used these results from \citere{paper} to define our 2-dimensional benchmark plane. 

Out of the eleven points provided in \citere{paper} we use eight points which all have $\vS > 30~\rm GeV$. These points 
are shown in \refta{musolf}.%
\footnote{We have checked that these eight points also lead to a FOEWPT using the more updated calculation
implemented in \texttt{BSMPTv3}~\cite{Basler:2018cwe,Basler:2020nrq,Basler:2024aaf}.}%
~They approximately fulfill the numerical conditions%
\footnote{For the four points with $\vS < 30 \gev$ the deviations from our three conditions exceed the few-percent level.} 
\begin{align}
    a_1\,\vS &= -32000\;,\\
    \lambda &= 0.18\;,\\
    b_3 &= -560\sqrt{b_4}\;, 
    \label{plane2con}
\end{align}
leaving the singlet vev $\vS$ and the parameter $b_4$ as our free parameters, so that we arrive at a 
two-dimensional benchmark plane that features a SFOEWPT.
The appearance of ``restricted allowed intervals'' for the Lagrangian parameters that can be observed in 
\refta{musolf} is just a consequence of the  requirement of a SFOEWPT as demanded in \citere{paper}.

\begin{table}[h]
\centering
\begin{tabular}{@{}ccccccccc@{}}
\toprule
Benchmark & $\vS$ [GeV]    & $\lambda$ & $a_1\rm~[GeV]$ & $a_2 $ & $b_3\rm~[GeV]$ & $b_4$ & $\kala$ & $\lahhH$ \\ \midrule
B1        & 60.9 & 0.17      & -490  & 2.65  & -361  & 0.52  & 1.42            & 0.25            \\
B2        & 59.6 & 0.17      & -568  & 3.26  & -397  & 0.78  & 1.40            & 0.31            \\
B3        & 54.6 & 0.17      & -642  & 3.80  & -214  & 0.16  & 1.41            & 0.34            \\
B4        & 47.4 & 0.18      & -707  & 4.63  & -607  & 0.85  & 1.47            & 0.38            \\
B5        & 40.7 & 0.18      & -744  & 5.17  & -618  & 0.82  & 1.47            & 0.37            \\
B6        & 40.5 & 0.19      & -844  & 5.85  & -151  & 0.08  & 1.48            & 0.42             \\
B7        & 36.4 & 0.18      & -898  & 7.36  & -424  & 0.28  & 1.43            & 0.48             \\
B8        & 32.9 & 0.17      & -976  & 8.98  & -542  & 0.53  & 1.41            & 0.54             \\ \bottomrule
\end{tabular}
\caption{Values of $\vS$, $\lambda$, $a_1$, $a_2$, $b_3$, $b_4$, $\kala$, and $\lahhH$ of the reference 
benchmark points from \protect\citere{paper}. }
\label{musolf}
\end{table}

\begin{figure}[htb]
    \centering
    \includegraphics[width=0.49\linewidth]{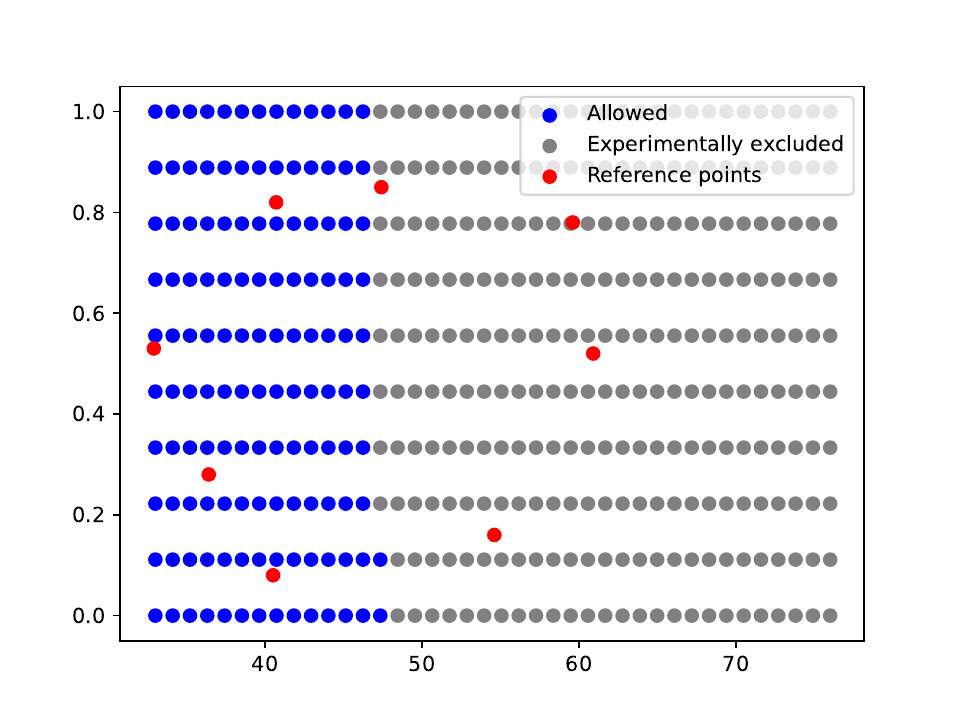}
    \includegraphics[width=0.49\linewidth]{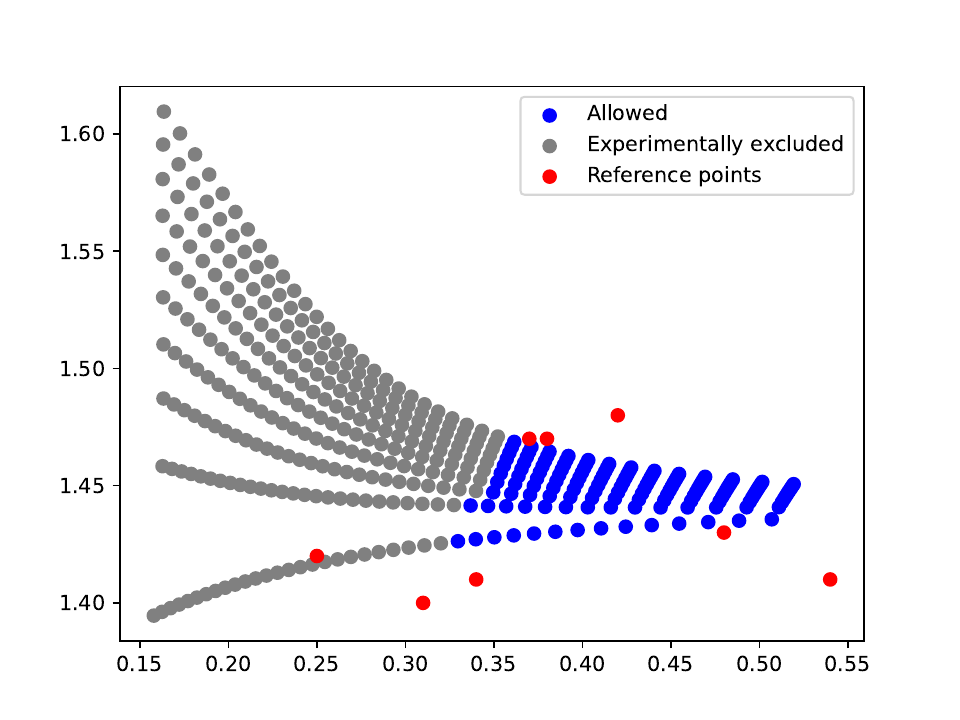}
    \includegraphics[width=0.49\linewidth]{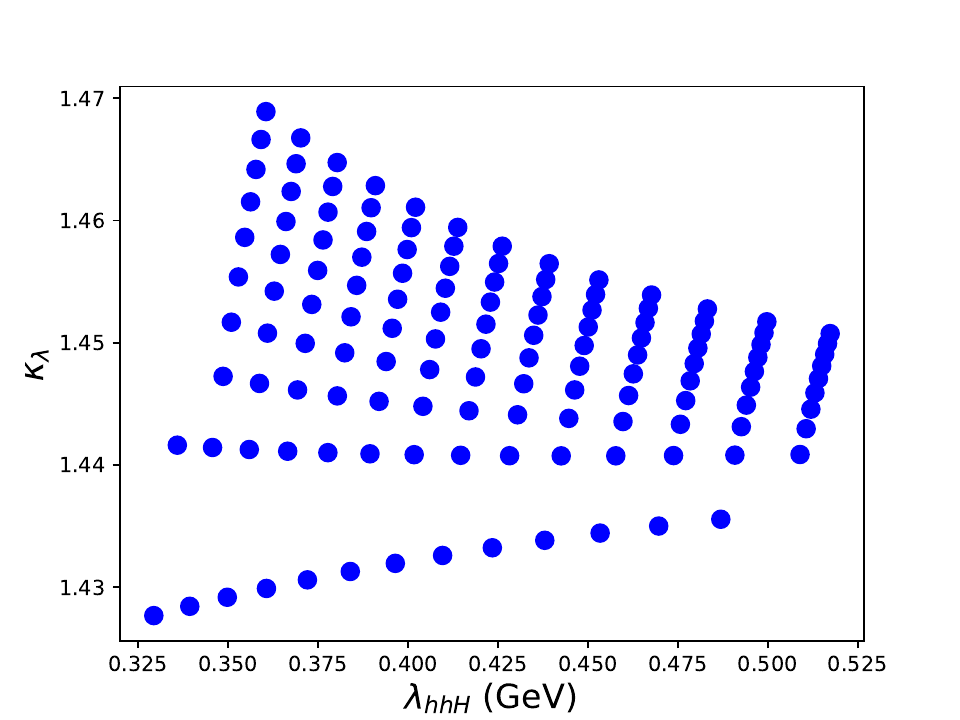}
    \caption{The benchmark plane with the experimentally excluded region in grey, the allowed region in blue and the points
    used to define the plane in red (see text). Upper left: the prediction in the plane $\vS$-$b_4$. 
    Upper right: the projection of the benchmark plane in the $\lahhH$-$\kala$ plane. 
    Lower plot: the final allowed benchmark plane in the $\lahhH$-$\kala$ projection.}
    \label{plane2}
\end{figure}

All points of the defined benchmark plane are within the range of parameters that were found to feature
a FOEWPT, see our analysis below.
Calculating the values of $a_1$, $\lambda$ and $b_3$ from the values of $b_4$ and $x$ in \refta{musolf}
we find a maximum deviation of $7\%$ for $a_1$, $6\%$ for $\lambda$ and $30\%$ for $b_3$ w.r.t.\ the values given in 
\refta{musolf}. 
We can use \refeq{mincon} and \refeqs{trans1} -- (\ref{trans3}) to translate the relations in \refeq{plane2con}
into the mass basis. 
This allows us to take into account the experimental and theoretical constraints as described in the previous subsection.
The plane is shown in the upper left plot of \reffi{plane2}, where the red stars indicate the eight benchmark points, 
and the blue points are allowed by all constraints.
The gray points are excluded by \texttt{HiggsBounds}, which found them to be incompatible with a 2020 ATLAS search for a
resonance of a heavy Higgs boson, which decays into two $Z$ bosons and ultimately into four leptons~\cite{ATLAS}. 
This experimental result was not yet available when \citere{paper} was published.
We also show the projection of the allowed/excluded regions (blue/grey) in the $\lahhH$-$\kala$ 
plane in the upper right plot of \reffi{plane2}. 
It can be seen how almost the entire plane lies in the area defined by the points of \cite{paper}, which ensures the 
SFOEWPT in our plane.
One can observe that four reference points are outside the plane, because of their smaller or larger
value of $\kala$. However, this does not constitute a problem, since the points are outside our benchmark plane by a 
maximum of $2\%$ in $\kala$. (In \citere{paper} a favorable value of $\kala \sim 1.5$ is found to ensure the SFOEWPT,
in agreement with our benchmark plane.)
Finally, in the lower plot of \reffi{plane2} we show a zoom into the allowed regions in the $\lahhH$-$\kala$ plane. 

\begin{figure}[htb]
    \centering
    \includegraphics[width=0.49\linewidth]{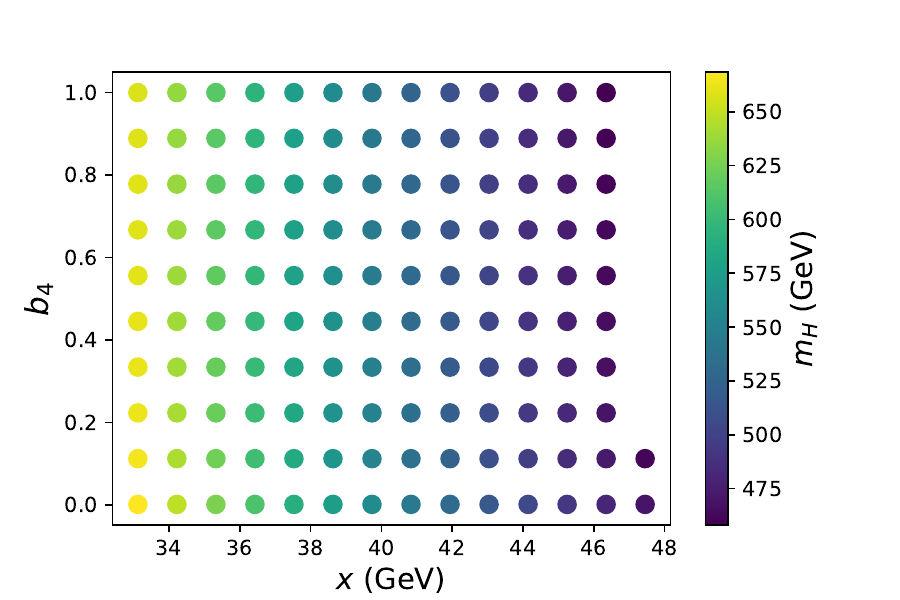}
    \includegraphics[width=0.49\linewidth]{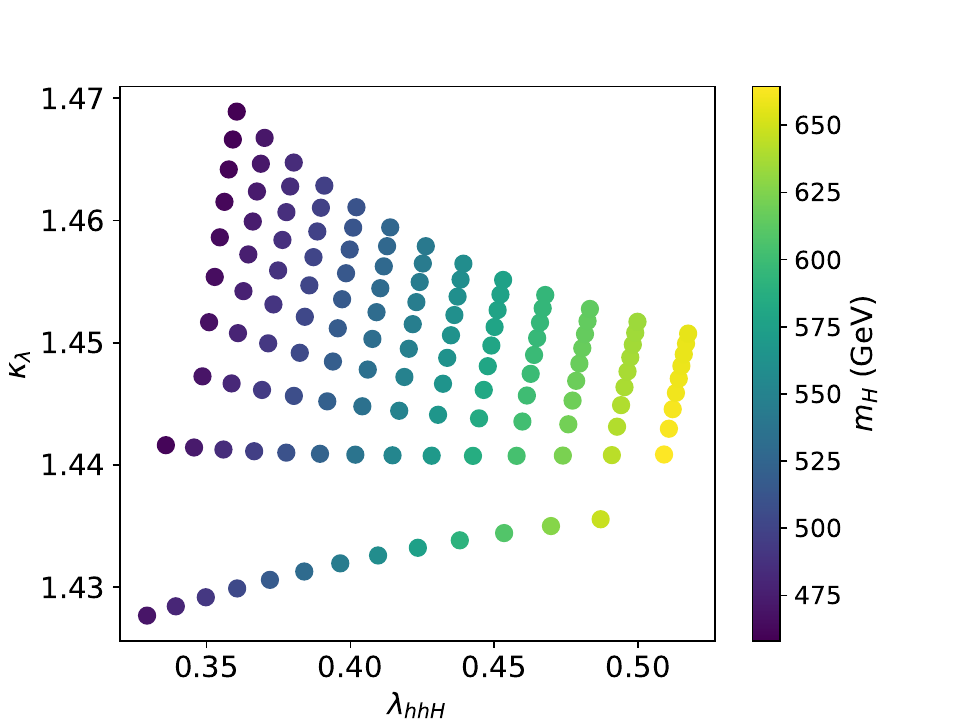}
    \caption{The prediction of the heavy Higgs mass $\MH$ in our benchmark scenario. Left: in the $\vS$-$b_4$ plane, right: in the $\lahhH$-$\kala$ plane.\label{m2}}
\end{figure}

\smallskip
In the following we briefly analyze the basic phenomenological features of our benchmark plane. 
In \reffi{m2} we show the prediction for $\MH$ in the $\vS$-$b_4$ plane (left) and in the $\lahhH$-$\kala$ plane (right). 
It can be seen that the mass of the heavy Higgs boson is inversely proportional to the singlet vev. 
This is an effect of the definition of the benchmark plane, not a feature of the model. The allowed range of $\MH$ in our
plane is $[458,660]~\rm GeV$. 
One can also see that $\lahhH$ increases with the heavy Higgs-boson mass.
Overall, we find that in our benchmark plane 
the product of $(\sin\theta \cdot \lahhH)$ is approximately constant (within a few percent). 
In the calculation of $gg \to H \to hh$ in the limit of large $\MH$ one thus finds a suppression 
with increasing $\MH$, and
in the limit $\MH \to \infty$ the amplitude for this process goes to zero. 

\begin{figure}[htb]
    \centering
    \includegraphics[width=0.49\linewidth]{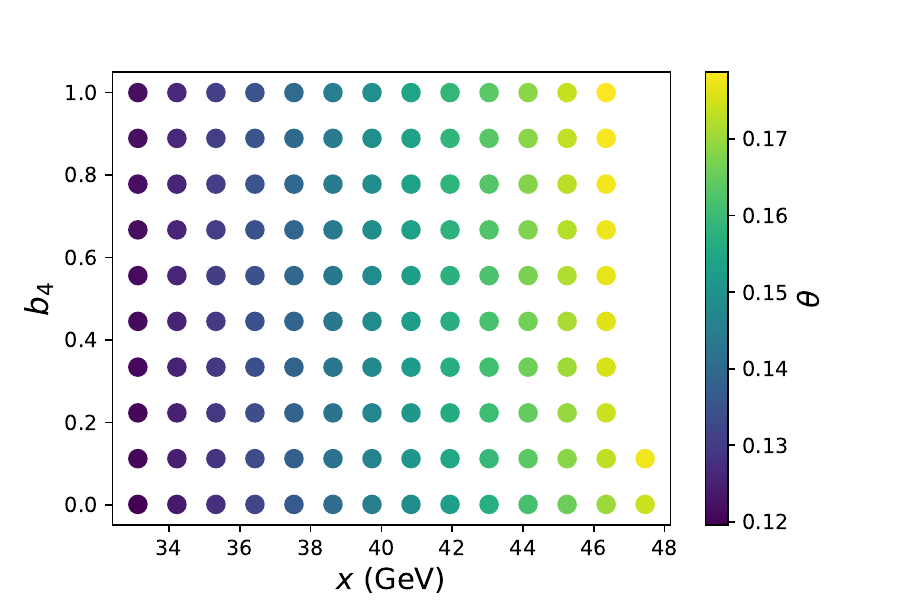}
    \includegraphics[width=0.49\linewidth]{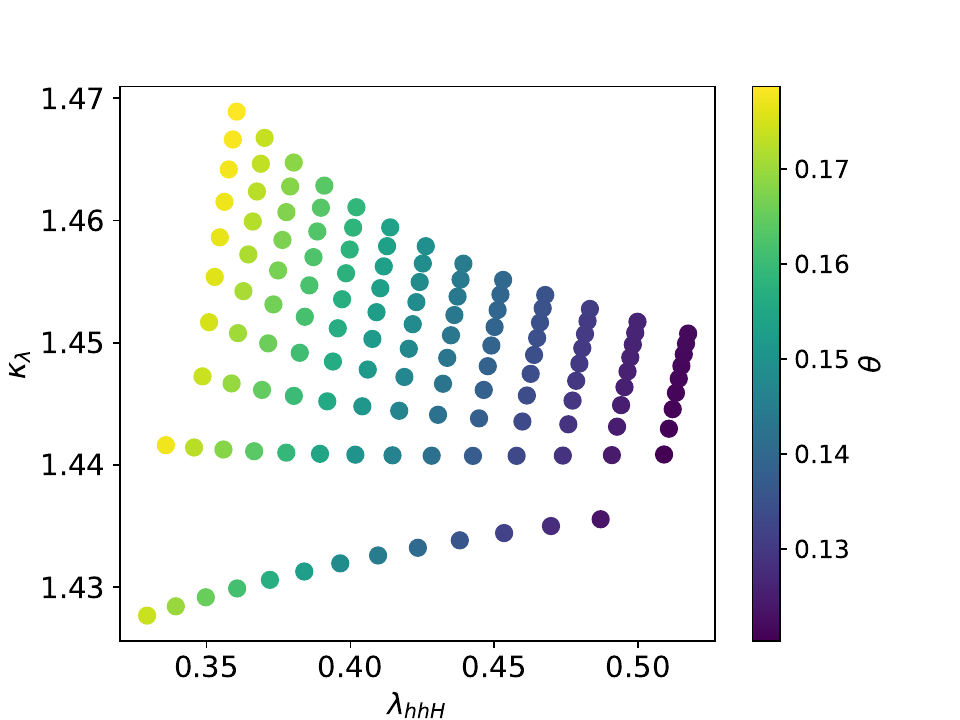}
    \caption{The prediction of the mixing angle in the second benchmark plane. Left: in the $\vS-b_4$ plane. Right: in the $\lahhH-\kala$ plane.\label{theta}}
\end{figure}

In \reffi{theta} we show  the prediction of the mixing angle $\theta$ in the $\vS$-$b_4$ plane (left) and in the 
$\lahhH$-$\kala$ plane (right). The cosine tends to 1 for small values of $x$, but $\cos\theta = 1$ is never reached,
which is again an artifact of our benchmark plane.
This is consistent since in the alignment limit the SM is recovered, in which no SFOEWPT is found, whereas it is ensured 
in our benchmark plane by construction. Similarly, $\kala = 1$ is never reached by construction. 

\clearpage

\section{The HL-LHC Analysis}
\label{sec:hllhc}

In this section, we first briefly review our set-up for the calculation of $pp \to hh$. We show the results for 
the cross section in our benchmark plane and analyze whether differences w.r.t.\ the SM predictions will be observable.
Finally, we analyze the sensitivity to the heavy Higgs-boson resonance and its THC, $\lahhH$. The corresponding analysis for 
high-energy $e^+e^-$ colliders is  presented in the subsequent \refse{sec:ilc}. 


\subsection{Calculation of \boldmath{$gg \to hh$}}
\label{sec:gghh}

\begin{figure}[h]
    \centering
    \includegraphics[width=0.49\linewidth]{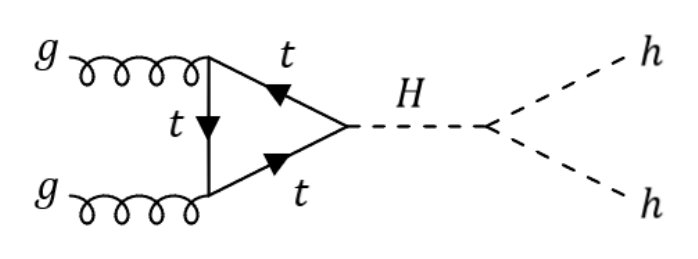}
    \includegraphics[width=0.49\linewidth]{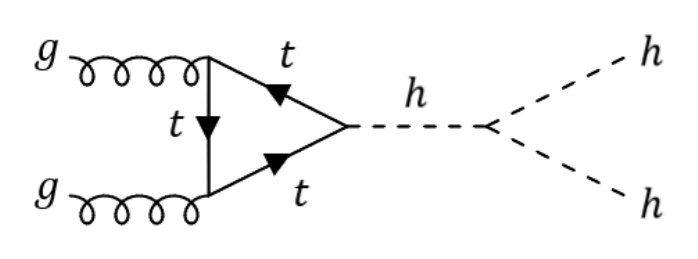}
    \includegraphics[width=0.42\linewidth]{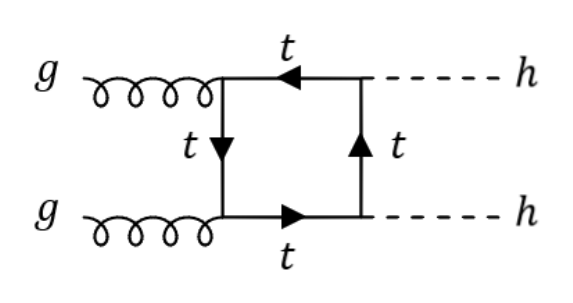}
    \caption{Feynman diagrams entering into the di-Higgs production cross section calculation in the RxSM. 
    Upper left: heavy triangle diagram; upper right: light triangle diagram; down: box diagram.}
    \label{diagrams}
\end{figure}

\noindent
Standard Model Higgs-pair production has not been observed yet at the LHC, and therefore di-Higgs
production constitutes an interesting window
to test new physics in future experiments such as the HL-LHC. The main production channel at the (HL-)LHC is gluon fusion.
In the SM at leading order (LO), two diagrams contribute to the process: the box diagram (shown by the lower diagram
in \reffi{diagrams}), which is given by a heavy quark loop with two Yukawa couplings, and the triangle diagram (shown by the upper
right diagram in \reffi{diagrams}), which is given by a heavy quark loop with a Yukawa coupling and the THC $\lahhh$.%
\footnote{We only display top-quark loops as in the SM the contributions from bottom-loops are very small.}
In the RxSM, an additional diagram contributes: the triangle diagram in which a heavy Higgs boson is propagating in the 
$s$-channel (shown by the upper left diagram in \reffi{diagrams}). It depends on the Yukawa coupling of the 
heavy Higgs boson and on the BSM THC $\lahhH$. The box and the light Higgs triangle diagram are known as 
the ``continuum'' or the non-resonant
part, while the heavy Higgs triangle diagram is called the ``resonant'' part (even in the case of $\MH < 2 \Mh$). 
It should be noted that in the SM there is a destructive interference between the triangle and the box diagram.

For our analysis we calculate the total production cross section as well as the
differential cross section with respect to the invariant mass of the two light Higgs bosons, \mhh.
For both calculations the code \texttt{HPAIR}~\cite{Dawson:1998py} has been used, adapted to the RxSM.
The original {\tt Fortran} code \texttt{HPAIR}  was written to calculate the production cross sections of two neutral Higgs bosons 
through gluon fusion in the SM and in the Minimal Supersymmetric Extension of the SM (MSSM). In the meantime, it has been extended 
to other models \citeres{Abouabid:2021yvw,Arco:2022lai,Dawson:1998py,Nhung:2013lpa,HP3,Grober:2015cwa}. 
For our project, we extended the code to include the RxSM.
The computation can be performed either at LO or next-to-leading order (NLO) in QCD. 
In the LO case, the calculation takes into account the top and bottom quark loops with full mass dependence, 
whereas in the NLO case the heavy top quark mass limit (HTL) is used. In this limit, the contributions of all quarks
are neglected, except for the top quark, which is assumed to be infinitely heavy. 
The computation of the NLO QCD corrections in the HTL including the full top-quark mass effects at LO leads to a $K$-factor,
i.e.~the ratio of the NLO over the LO cross section, of approximately 2 in the SM \cite{Dawson:1998py} and also in
other BSM extensions,
as found in \cite{Abouabid:2021yvw,Arco:2022lai,Dawson:1998py,Nhung:2013lpa,HP3,Grober:2015cwa}.%
\footnote{In our calculations performed for this work we confirmed a $K$-factor of $\sim 2$ for the RxSM, see the next subsection.}
This value approximates the results of the 
$K$-factor of the inclusive cross section including the finite top-mass effects at NLO QCD, very well as shown 
in \citeres{Borowka:2016ehy,Borowka:2016ypz,Baglio:2018lrj,Baglio:2020ini,Baglio:2020wgt} for the SM and \cite{mul,nlothdm} for the 2HDM.
The QCD corrections are not affected by Higgs mixings, so they can be taken over to the RxSM case. 

The code also  provides the calculation of differential cross sections with respect to the invariant Higgs-pair mass, \mhh,
both at LO and in the HTL. However, as has been shown in \citere{Buchalla:2018yce} in the context of non-linear 
effective field theory, 
mass effects can be significant in the \mhh\ distributions. Since for BSM models there are no results available for
distributions at NLO QCD including the full mass dependence, we will stick to LO distributions in the following. 
Although we are aware that NLO
corrections are important, we do not want to present results for distributions that could be significantly distorted by mass effects, and hence chose to make this compromise.
Since not all calculations necessary for our analysis can be performed at NLO, we consistently use LO QCD everywhere.


\subsection{Analysis of the Cross Section Predictions}
\label{sec:xs-anal}

\begin{figure}[htb]
    \centering
    \includegraphics[width=0.49\linewidth]{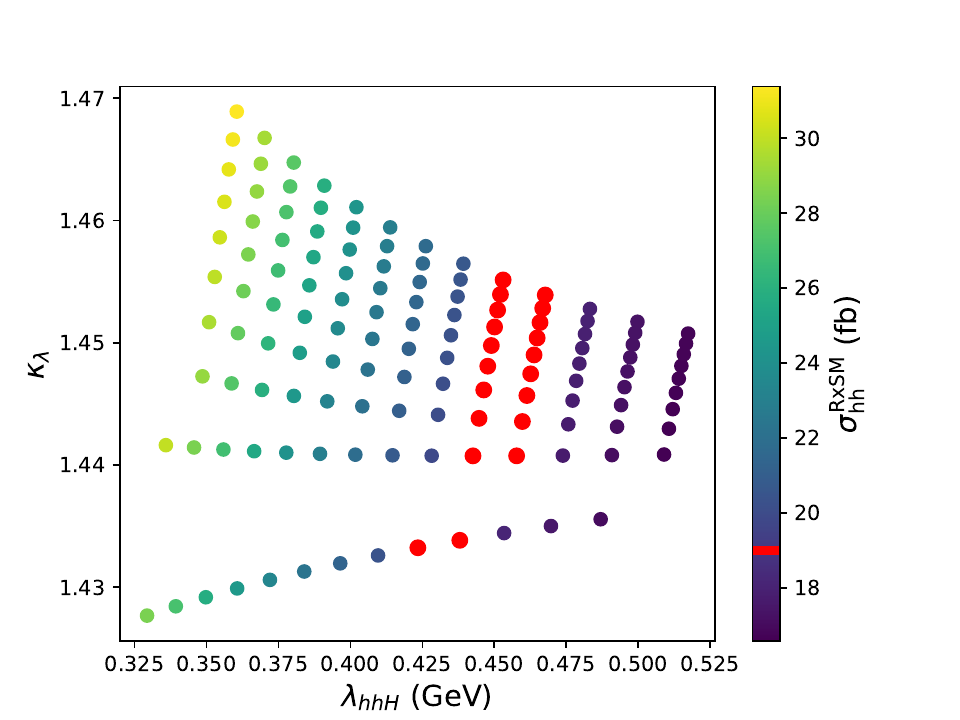}
    \includegraphics[width=0.49\linewidth]{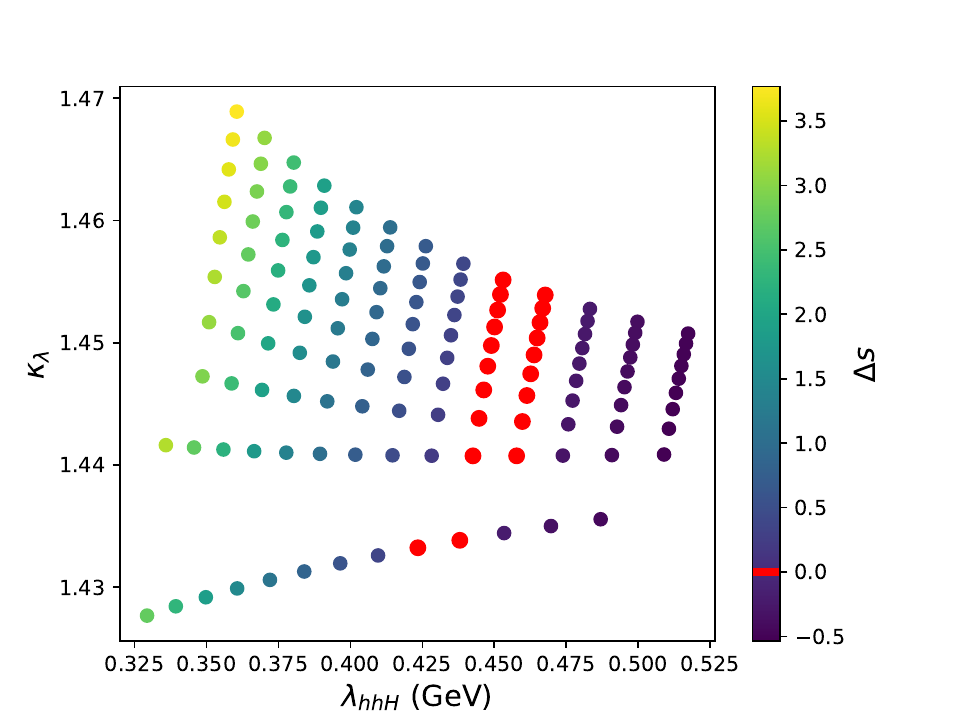}
    \caption{The $\lahhH$-$\kala$ benchmark plane. 
Left: total cross section of the process $gg \to hh$ for the HL-LHC in the RxSM. The red mark on the 
colorbar corresponds to $\sigRx = \sigSM$. The red colored points are defined via the right plot (see text).
Right: the significance of the cross section of the process $gg \to hh$ in the RxSM w.r.t.\ the SM. 
The red mark on the colorbar corresponds to $\Delta s = 0$, and the red points have $|\Delta s| < 0.1$.}
    \label{cross}
\end{figure}

In \reffi{cross} we show the results for the
di-Higgs production cross section in the RxSM at LO. 
We have verified numerically that the results at NLO QCD are larger by a factor $\sim 2$. 
From now on, we concentrate on the LO result (keeping in mind the NLO factor). 
All cross section calculations are done for $\sqrt{s} = 14 \tev$.
The corresponding SM cross section is
obtained with our RxSM version of \texttt{HPAIR} setting the mixing angle $\theta = 0$, i.e.\ in the SM limit.
Numerically, we find 
\begin{equation}
    \sig^{\rm LO}_{\rm SM}\; (pp \to hh) = 19.76\; {\rm fb}\,, \quad
    \sig^{\rm NLO}_{\rm SM}\; (pp \to hh) = 38.24\; {\rm fb}\,.
    \label{eq:smvalues}
\end{equation}

In a first step, we compare the results of the RxSM with the SM model ones at the HL-LHC and analyze whether the cross sections
of both models can be distinguished experimentally. 
To do so we define the statistical significance of the RxSM cross section w.r.t.\ the SM.  
To calculate the uncertainty of the cross section measurement at the HL-LHC we take the anticipated significance of 
the SM di-Higgs production cross section at the HL-LHC~\cite{Cepeda:2019klc}, which has been found to be $s_{\rm SM}=4.5\sigma$.
This significance is rescaled to the cross section of the RxSM (as the number of expected events will be different).
Since we are dealing with a Gaussian distribution, the uncertainty scales as the square root of the number of events, 
or in this case of the cross section. The uncertainty $\Delta\sigRx$ on the cross section in the RxSM can then be 
obtained as,%
\footnote{Since, as discussed above, also in the RxSM the $K$-factor is $\sim 2$ as in the SM, 
we can safely assume that the NLO QCD
corrections, that are not taken into account (see the discussion above), rescale all cross
sections and the corresponding uncertainties in the same way.}%
\begin{equation}
    \Delta\sigRx=\frac{\sigSM}{s_{\rm SM}}\sqrt{\frac{\sigSM}{\sigRx}},
\end{equation}
and the significance of the deviation of the total RxSM di-Higgs production from the SM prediction is defined as,
\begin{equation}
    \Delta s=\frac{\sigRx-\sigSM}{\Delta\sigRx}\,.
    \label{significance}
\end{equation}

In \reffi{cross} we show the $\lahhH$-$\kala$ benchmark plane as defined in the previous section. 
In the left plot we display as color coding the total cross section of the di-Higgs production in the RxSM and we mark in red when the value of
the cross-section in the RxSM is approximately equal to the SM, which we define through the right plot, where 
we show as color coding the significance of the deviation from the SM, i.e.\ $\Delta s$ as defined in  
\refeq{significance}, where red points have $|\Delta s < 0.1|$.
For values of $\lahhH \lsim 0.4255$ we find larger RxSM cross section values than in the SM, whereas for larger $\lahhH$ values
they are smaller. However, this is not a general feature, but an artifact of our benchmark plane.
The figure shows that the cross-section of the process is inversely proportional to the THC $\lahhH$. Here it is 
important to keep in mind that $\lahhH$ increases with $\MH$. However, as argued in the section where the benchmark plane
is defined, an overall suppression of the heavy Higgs-boson contribution is expected with increasing $\lahhH$ due to the fact
that in our benchmark plane we find that the product of $(\sin\theta \cdot \lahhH)$ is approximately constant.
Finally, as can be observed in the right plot of \reffi{cross}, 
we find that for the smallest allowed values of $\lahhH$ in our benchmark plane the cross section of the RxSM
deviates by more than $3\,\sig$ from the SM prediction, i.e.\ from the cross section measurement alone a difference 
could be observed. For most parts of the parameter space, however, this difference is too small to be significant.

In the region where $\sigSM \approx \sigRx$, i.e.\ the region marked in red in the right plot 
of \reffi{cross}, the RxSM is not in the alignment limit, but the various BSM effects cancel each other.
W.r.t.\ the SM one has $\kala \sim 1.45$, i.e.\ the destructive interference of the box diagram and the $h$ $s$-channel
contribution is enhanced, leading to a smaller cross section. This, however, is compensated by the 
resonant $H$ exchange contribution, leading to an accidental numerical cancellation of both effects.

\begin{figure}[h]
    \centering
    \includegraphics[width=0.8\linewidth]{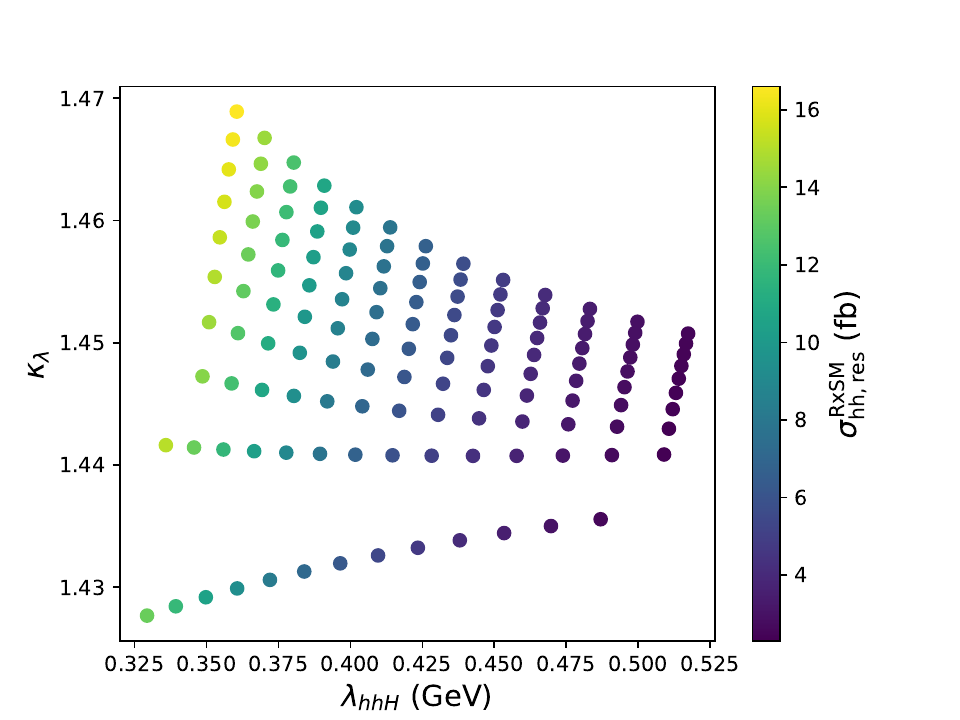}
    \caption{The cross section of the process $gg \to hh$ for the HL-LHC in the RxSM only taking into account 
    the heavy Higgs boson triangle diagram, shown in the plane $\lahhH$-$\kala$.}
    \label{cross_resonly}
\end{figure}

Resonant di-Higgs-boson searches at ATLAS \citeres{ATLAS:2022kbf,ATLAStrilinear} and CMS \citeres{CMS:2022dwd,CMStrilinear} so far take into account only the resonant diagram, but neglect possible effects from
the two continuum diagrams, which will be discussed further in \refse{sec:mhh-hllhc}. 
Here, in 
\reffi{cross_resonly}, we show the total cross section in the RxSM for the case that the continuum diagrams are 
(incorrectly) not taken into account. It can be observed that this cross section is lower than the SM result by $\sim 20\%$
for small values of $\lahhH$ and by up to $\sim 80\%$ for large values of $\lahhH$. These numbers differ substantially from the 
complete calculation taking into account all diagrams, as shown in \reffi{cross}. This demonstrates already at the level
of the full cross section that the experimental ``approximation'' of neglecting the continuum diagrams may not be adequate 
in all cases.

\subsection{Analysis of \boldmath{\mhh}}
\label{sec:mhh-hllhc}

\subsubsection{Definitions and Benchmark Points}

\begin{figure}[htb]
    \centering
    \includegraphics[width=0.7\linewidth]{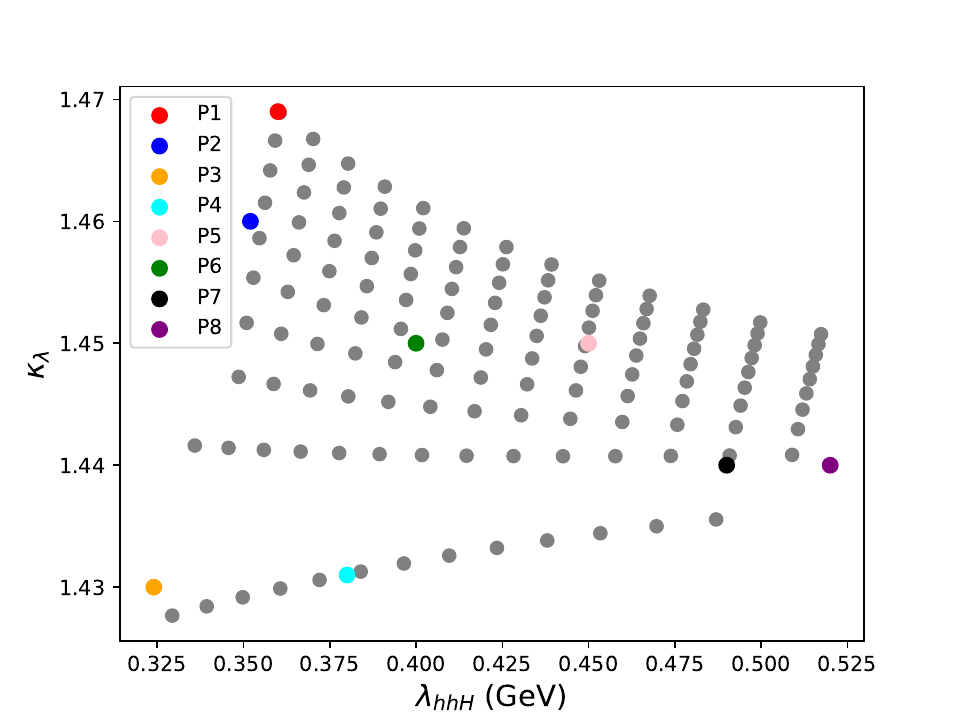}
    \caption{The $\lahhH$-$\kala$ plane with the eight benchmark points marked. For details, see \protect\refta{points2}.}
    \label{points3}
\end{figure}

Experimental di-Higgs analyses not only rely on the total cross section, but also build substantially on 
differential distributions like the differential invariant mass distribution of the di-Higgs system, \mhh\ 
(which we evaluate, as discussed above, also with the code \texttt{HPAIR}).
This will be particularly relevant in parameter regions where $\Delta s < 3$, i.e.\ the measurement of the 
total cross section is not sufficient to distinguish the RxSM from the SM.
To facilitate our analysis, we are going to focus on eight benchmark points distributed over the plane. 
They have been selected to explore parameter regions with different values of the BSM Higgs-boson mass, 
different couplings and also different values of the statistical significance of the total cross section 
with respect to the SM. The eight benchmark points that we have defined are shown in \reffi{points3}, 
with their input parameters and other relevant quantities
given in \refta{points2}. In our analysis we will group these benchmark points according 
to the differences between $\sigRx$ and $\sigSM$ in each point. Specifically, we define 
\begin{itemize}
    \item {\bf Region 1:} $\Delta s > 3$\,,
    \item {\bf Region 2:} $3 > \Delta s > 0.5$\,,
    \item {\bf Region 3:} $0.5 > \Delta s > -0.5$\,.
\end{itemize}

\begin{table}[htb!]
\centering
\begin{tabular}{@{}ccccccccccc@{}}
\toprule
Id & $m_H~\rm [GeV]$ & $\theta$ &$\vS\rm~[GeV]$ &$b_4$ & $\lahhH$ & $\kala$ & $\Gamma_{H}~\rm [GeV]$ & 
$\sigma_{hh}^{\rm RxSM}~\rm [fb]$ &$\Delta s$& $R$  \\ \midrule
P1 & 459.2         & 0.178 & 46.3 & 0.89    & 0.36           & 1.47               & 3.81         
   & 31.1           & 3.7           & 235    \\
P2 & 464.9         & 0.176  & 46.3 & 0.45   & 0.35           & 1.46              & 3.73         
   & 30.0            & 3.2                & 224   \\
P3 & 469.4         & 0.174 & 47.4 & 0.00    & 0.32           & 1.43               & 3.48         
   & 28.4              & 2.7               & 206 \\
P4 & 529.8        & 0.153  & 41.9 & 0.00    & 0.38          & 1.43              & 4.15         
   & 23.3                     & 1.1        & 158 \\
P5 & 577.5         & 0.139 & 37.5 & 0.78    & 0.45          & 1.45              & 5.05         
   & 19.5               & 0.1              & 111   \\
P6 & 531.7        & 0.152  & 40.8 & 0.45   & 0.40         & 1.45              & 4.49         
   & 22.4                  & 0.9           & 146  \\
P7 & 642.9         & 0.125 & 34.2 & 0.11    & 0.49          & 1.44               & 5.5         
   & 17.5                & -0.3             & 80    \\
P8 & 657.9         & 0.122 & 33.1 & 0.78    & 0.52          & 1.44              & 5.84         
   & 17.3                   & -0.4         & 76  \\
\bottomrule
\end{tabular}
\caption{Benchmark points: identifier, heavy Higgs mass, mixing angle, singlet vev, $b_4$, $\lahhH$ coupling, 
$\kala$ modifier, heavy Higgs decay width, di-Higgs production cross section at the HL-LHC, $\Delta s$, 
and the $R$ parameter defined in \refeq{rpar}, see text.}
\label{points2}
\end{table}

\begin{figure}[h]
    \centering
    \includegraphics[width=0.8\linewidth]{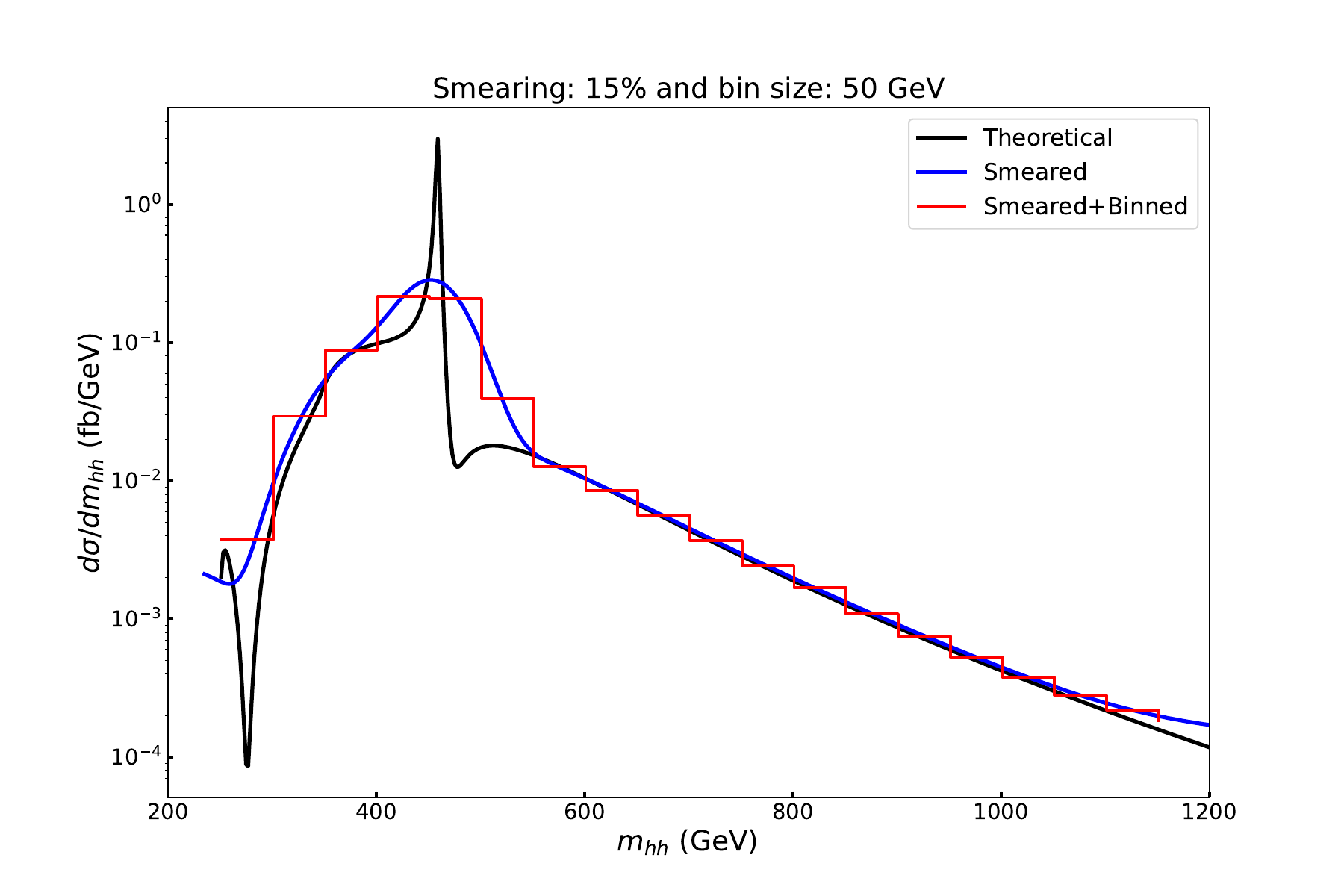}
    \caption{Differential cross section of the process $pp\to hh$ at the HL-LHC with $\sqrt{s} = 14 \gev$ 
    as a function of \mhh\ for the point P1, see \protect\refta{points2}; 
    black: theoretical curve, blue: smeared curve with 15\% of smearing, 
    red: smeared and binned curve with 15\% of smearing and a bin size of $50 \gev$.}
    \label{smearing}
\end{figure}

Before we present our \mhh\ analysis, we discuss the impact of the experimental uncertainties, see \citere{Arco:2022lai}.
In \reffi{smearing} we show the theoretical prediction for the \mhh\ distribution in black for the benchmark point P1,
as given in \refta{points2}. It has a dip for low values of the invariant mass, $\mhh \sim 290 \gev$. 
This is caused by a negative interference between the box and the light Higgs-boson triangle diagrams. 
In the SM with $\kala \sim 1$ this interference occurs at $\mhh \sim 250 \gev$. In P1 we have $\kala \sim 1.5$,
and the negative interference is shifted to $\mhh \sim 290 \gev$. 
The second important effect is the peak-dip structure observed around the resonance of the heavy Higgs boson,
$\mhh = \MH \sim 460 \gev$. This peak-dip structure is due to the interference of the heavy Higgs-boson triangle
and the two non-resonant diagrams, see the discussion in \citere{Arco:2022lai} and references therein.
The sign of the couplings entering the heavy Higgs-boson resonance diagram, $Y_t\cdot\lahhH$, where $Y_t$ is the top Yukawa 
coupling of the heavy Higgs boson, determines whether one finds a peak-dip structure (as in our case), or a dip-peak structure. 

The first experimental uncertainty that has to be taken into account is the uncertainty in the \mhh\ measurement,
labeled as ``smearing''. Here we follow the procedure of \citere{Arco:2022lai}, where each point in \mhh\ is smeared out
out as a Gaussian distribution in \mhh. We represent each point in \mhh\ as a Gaussian distribution with a 
full width half maximum (FWHM) of a percentage ($p$) of the corresponding value of \mhh, see \citere{Arco:2022lai} for details,
where it was argued that the percentage ($p$) value to perform a realistic analysis is $15\%$.
The effect of smearing on a distribution can be seen in the blue curve in \reffi{smearing}.
It can be observed how the distribution is smoothened out and does not exhibit a pronounced peak-dip structure anymore,
which will make it more difficult to identify the resonance contribution. 
The second effect to be taken into account is that the detector does not have an infinite resolution in \mhh. 
Instead, the data will be given in bins of $50 \gev$ width, see again the discussion in \citere{Arco:2022lai}. 
Taking this into account on top of the smearing results in the red curve shown in \reffi{smearing}. Indentifying the 
resonance contribution becomes even more difficult taking the finite resolution in \mhh\ into account.

One main objective of our analysis is the question whether we can distinguish the RxSM from the SM via the \mhh\ distributions, i.e.\ 
whether the effect of the heavy Higgs-boson resonance can be detected. To this end, we define a theoretical parameter, $R$, to compare quantitatively the difference
between the RxSM and the SM distributions for the different benchmark points.  Following \citere{Arco:2022lai}, we define $R$ as, 
\begin{equation}
    R=\frac{\sum_i|N_i^R-N_i^C|}{\sqrt{\sum_iN_i^C}}\,, 
    \label{rpar}
\end{equation}
where $N^R_i$ is the number of events of the RxSM distribution, and $N^C_i$ is the number of events of the SM distribution in bin~$i$. 
The window in which the bins are counted is defined by~\cite{Arco:2022lai}, 
\begin{equation}
    |N^R-N^C|>\rm bin~size\cdot20~\gev\,,
    \label{rpar-cond}
\end{equation}
i.e., the sum over $i$ in \refeq{rpar} runs over all the bins that fulfill the condition in \refeq{rpar-cond}. 
With this choice
we focus on the region in which there are sizeable differences between the two distributions, i.e. around the resonance. 
It is important to emphasize that $R$ is a theoretical measure to compare the two distributions with each other. 
To determine whether via an \mhh\ distribution the measurement the value of the THC \lahhH\ can be performed, a full 
experimental analysis is required, which is beyond the scope of this paper.


\subsubsection{Complete Calculation}

In this subsection we present our results for the calculation of the differential cross section based on the full set of LO
diagrams. In the next subsection we highlight the differences w.r.t.\ the calculation taking into account only the resonant
heavy Higgs-boson diagram, as done by the experimental collaborations to obtain their exclusion limits.

\subsubsection*{Region 1}

\begin{figure}[htb]
    \centering
    \includegraphics[width=0.8\linewidth]{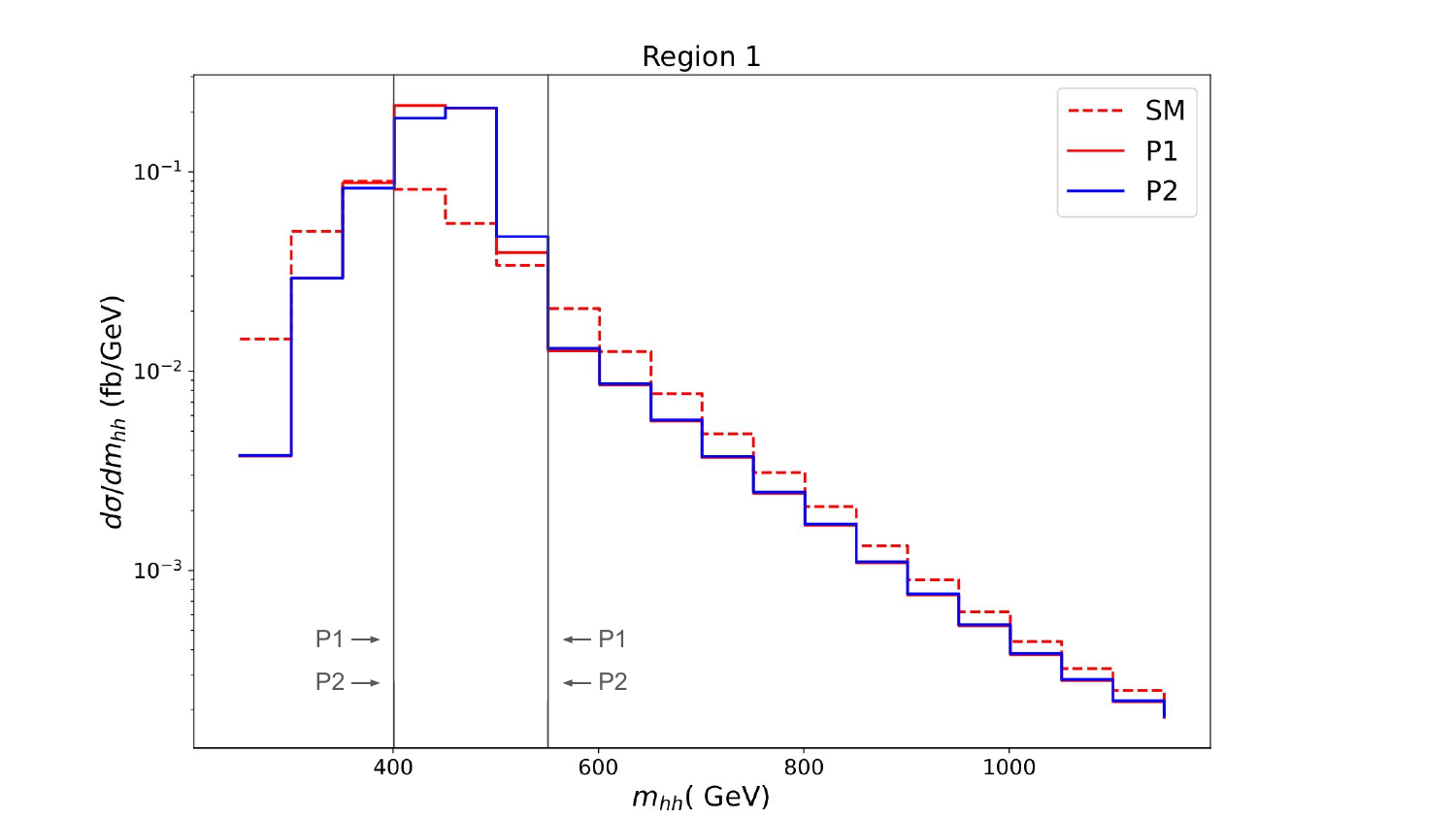}
    \caption{Differential cross section of the process $pp\to hh$ at the HL-LHC as a function of $m_{hh}$ for the SM (red dashed line) 
    and for two RxSM benchmark points: P1 (red) and P2 (blue), see \protect\refta{points2}. 
    Results are shown for a smearing of 15\% and a bin size of $50 \gev$. The bins that have been used to calculate the $R$ values 
    are indicated by gray vertical lines.} 
    \label{hist_12}
\end{figure}

In the first region with $\Delta s > 3$ one could observe indications of BSM physics via the total cross-section
measurement alone. Two benchmark points, P1 and P2, lie in Region~1. 
In \reffi{hist_12} we show the \mhh\ distributions for a smearing of 15\% and a bin size of $50 \gev$, as 
discussed in the previous subsection.
Compared are the results for the SM (red dashed line), benchmark point P1 (red) and P2 (blue). 
One can observe that the ``original'' dip-peak structure, as e.g.\ visible in \reffi{smearing}, 
is not visible anymore, an effect of the
smearing and binning. On the other hand, a pronounced peak w.r.t.\ the SM is visible around $\mhh \sim 460 \gev$, 
which corresponds to a 
good approximation to the values of $\MH$ in P1 and P2. The values found for $R$ according to \refeqs{rpar} and 
(\ref{rpar-cond}) are $R \sim 230$ (see also \refta{points2}), 
where we have indicated in \reffi{hist_12}, which bins are taken into account in the respective evaluation. 
While $R$ is not representing a true experimental significance, the values are relatively high, giving rise to the
expectation that the RxSM and the SM can be distinguished
not only via a measurement of the total cross section, but also via a measurement of the \mhh\ distributions.

\subsubsection*{Region 2}

\begin{figure}[htb]
    \centering
    \includegraphics[width=0.8\linewidth]{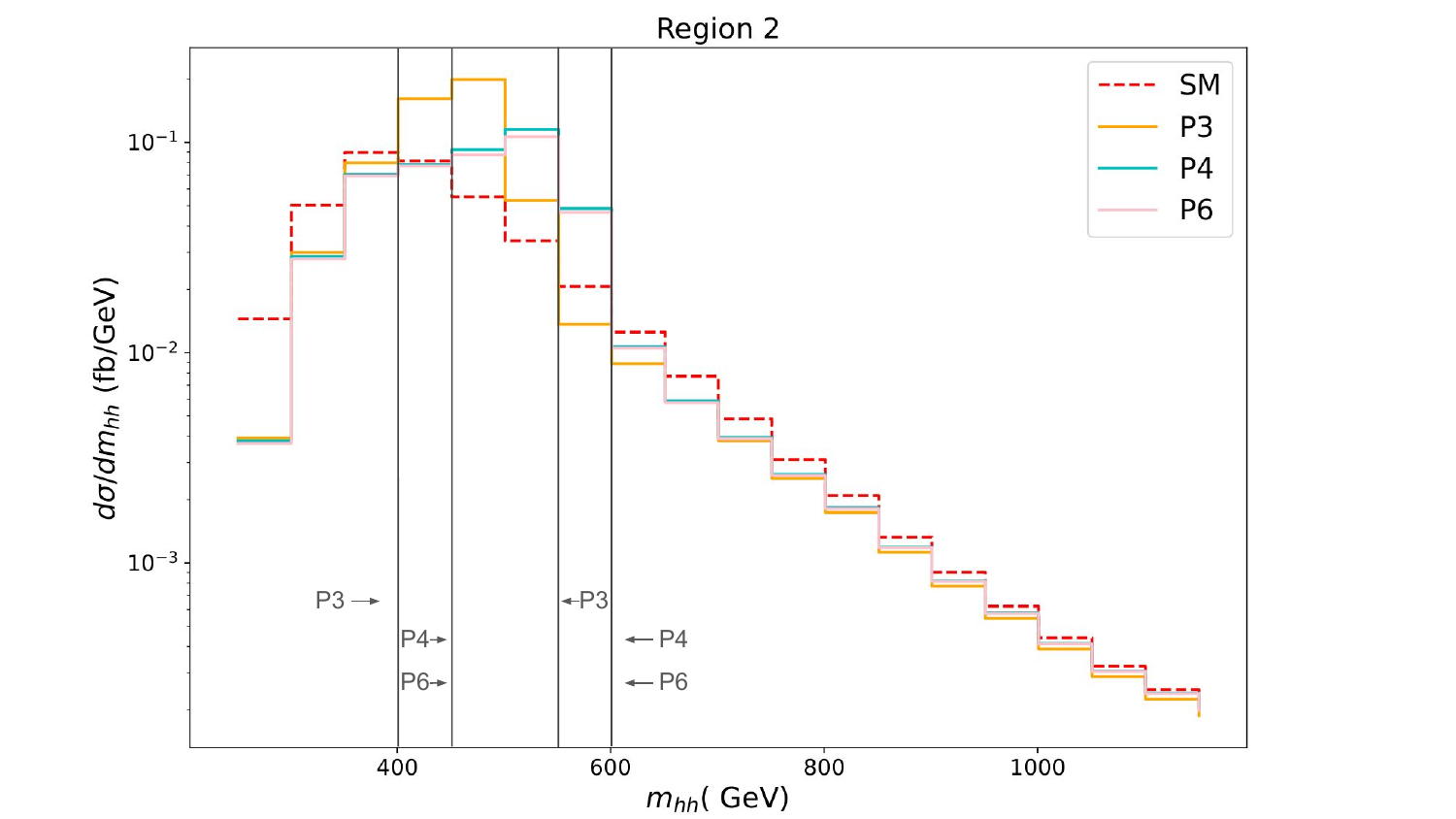}
    \caption{Differential cross section of the process $pp\to hh$ at the HL-LHC as a function of $m_{hh}$ for the SM (red dashed line) 
    and for three RxSM benchmark points in region 2: P3 (orange), P4 (cyan) and P6 (pink) , see \protect\refta{points2}. 
    Results are shown for a smearing of 15\% and a bin size of $50 \gev$. The bins that have been used to calculate the $R$ values are indicated with gray lines.
    }
    \label{hist_456}
\end{figure}

In the second region the difference in the total di-Higgs production cross section between the RxSM and the SM is $3 > \Delta s > 0.5$, i.e.\ 
the total cross section measurement would not exhibit a significant deviation. However, as discussed above, this is due to the cancelation 
of several BSM effects, as we will demonstrate here. In \reffi{hist_456} we show the \mhh\ distributions of the three benchmark points 
in region 2, P3 (orange), P4 (cyan) and P6 (pink), see \refta{points2}. As before, they are compared to the SM distribution (dashed red), and 
a smearing of 15\% and a  bin size of $50 \gev$ have been applied.
As for region 1, it can be observed how the peak-dip structure is washed out, leaving resonance peaks around the values of $\MH$, with 
$\MH \sim 470 \gev$ for P3 and $\MH \sim 530 \gev$ for P4 and P6 (see also \refta{points2}), 
where we have indicated in \reffi{hist_456}, which bins are taken into account in the respective evaluation. Most importantly, all three RxSM distributions differ substantially from the 
SM \mhh\ distribution where the differential cross section is large. 
The values of $R$ found for the three points are $\sim 200$ for P3, and $\sim 150$ for P4 and P6. These large values give 
rise to the hope that while the total cross section does not allow to distinguish the RxSM from the SM, such a distinction may be possible via the
measurement of the \mhh\ distribution in comparison with the theory prediction for the SM.

\subsubsection*{Region 3}

\begin{figure}[h]
    \centering
    \includegraphics[width=0.85\linewidth]{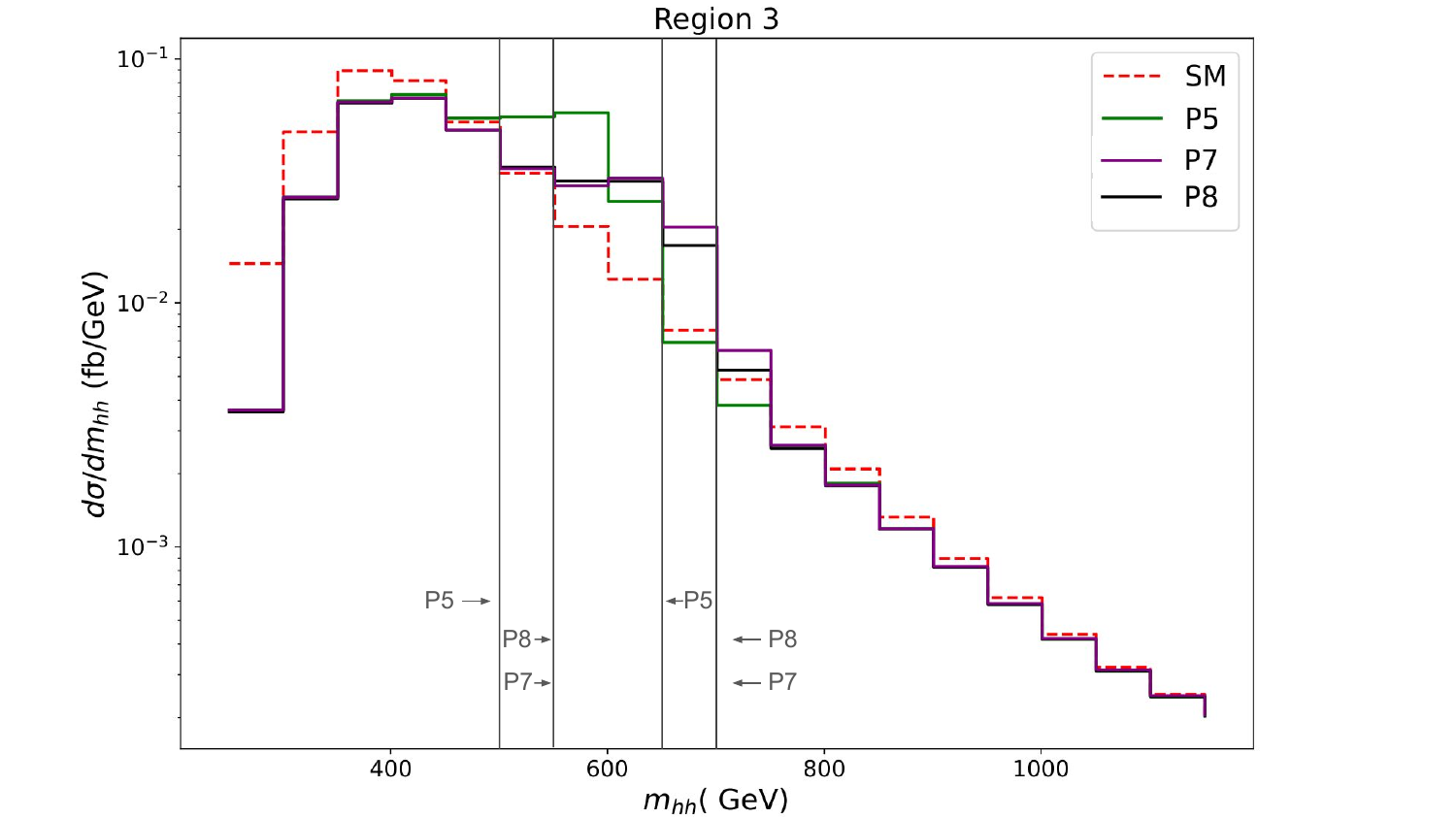}
    \caption{Differential cross section of the process $pp\to hh$ at the HL-LHC as a function of $m_{hh}$ for the SM 
    (red dashed line) 
    and for three RxSM benchmark points in region 3: P5 (green), P7 (purple) and P8 (black), see \protect\refta{points2}. 
    Results are shown for a smearing of 15\% and a bin size of $50 \gev$. The bins that have been used to calculate the
    $R$ values are indicated with gray lines. 
    }
    \label{hist_7}
\end{figure}

In the third region the difference in the total di-Higgs production cross section between the RxSM and the SM 
is $0.5 > \Delta s > -0.5$, i.e.\ the total cross section in the RxSM is effectively identical to the SM prediction. 
In \reffi{hist_7} we show the \mhh\ distributions of the three 
benchmark points in region 3, P5 (green), P7 (purple) and P8 (black), see \refta{points2}. As before, 
they are compared to the SM distribution 
(dashed red), and a smearing of 15\% and a  bin size of $50 \gev$ have been applied.
From the original peak-dip structure only a broadly smeared out ``resonance peak'' remains, again centered around the 
respective values of the heavy Higgs-boson mass. 
However, in contrast to regions~1 and~2, the differences w.r.t.\ the SM \mhh\ distribution around $\MH$, i.e.\ where 
the differential cross section is relatively large, appears much smaller than in the regions~1 and~2. Correspondingly,
relatively smaller $R$ values are found:
$R \sim 110$ for P5, and $R \sim 80$ for P7 and P8 (see also \refta{points2}), where again we have indicated in 
\reffi{hist_7}, 
which bins are taken into account in the respective evaluation. While these values still appear substantial, 
a more realistic experimental analysis
will be needed to determine whether in region~3 the \mhh\ measurement at the HL-LHC will be able to distinguish 
the RxSM from the SM.

\subsubsection{Pure Heavy Resonant Contribution}

In view of recent improvements in the experimental sensitivity to resonant di-Higgs
production (see, e.g., \citeres{CMStrilinear,ATLAStrilinear}) it is crucial that the experimental limits 
(and possibly eventually also the experimental
measurements) are presented in a way that they can be correctly confronted with theoretical
predictions in different models. The resonant limits that have been presented by ATLAS~\cite{ATLAStrilinear} 
and CMS~\cite{CMStrilinear}
so far were obtained assuming that only one heavy resonance is contributing to the cross section, 
neglecting the non-resonant contributions.
In \citere{Heinemeyer:2024hxa} it was demonstrated for benchmark points in the 2HDM that the current experimental 
procedure may not yield reliable resonant di-Higgs exclusion limits. In this subsection we compare the \mhh\ 
distributions of 
the full calculation, as presented in the previous subsection, with the distributions obtained neglecting the non-resonant 
contributions, i.e.\ \mhh\ distributions of the type employed by the experimental collaborations to obtain their current
exclusion bounds. The comparison is shown for one benchmark point in each region. 

\begin{figure}[htb!]
    \centering
    \includegraphics[width=0.6\linewidth]{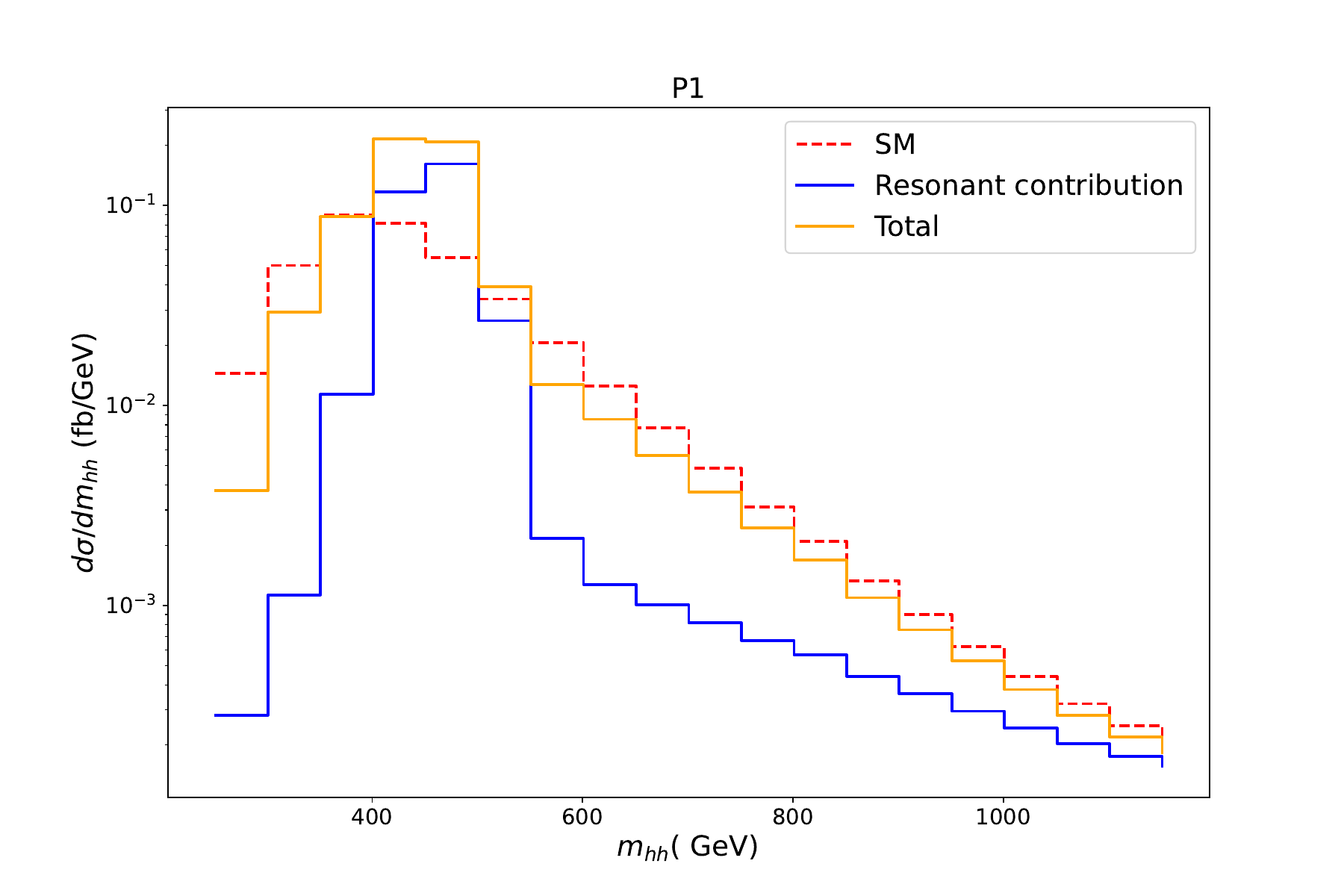}
    \includegraphics[width=0.6\linewidth]{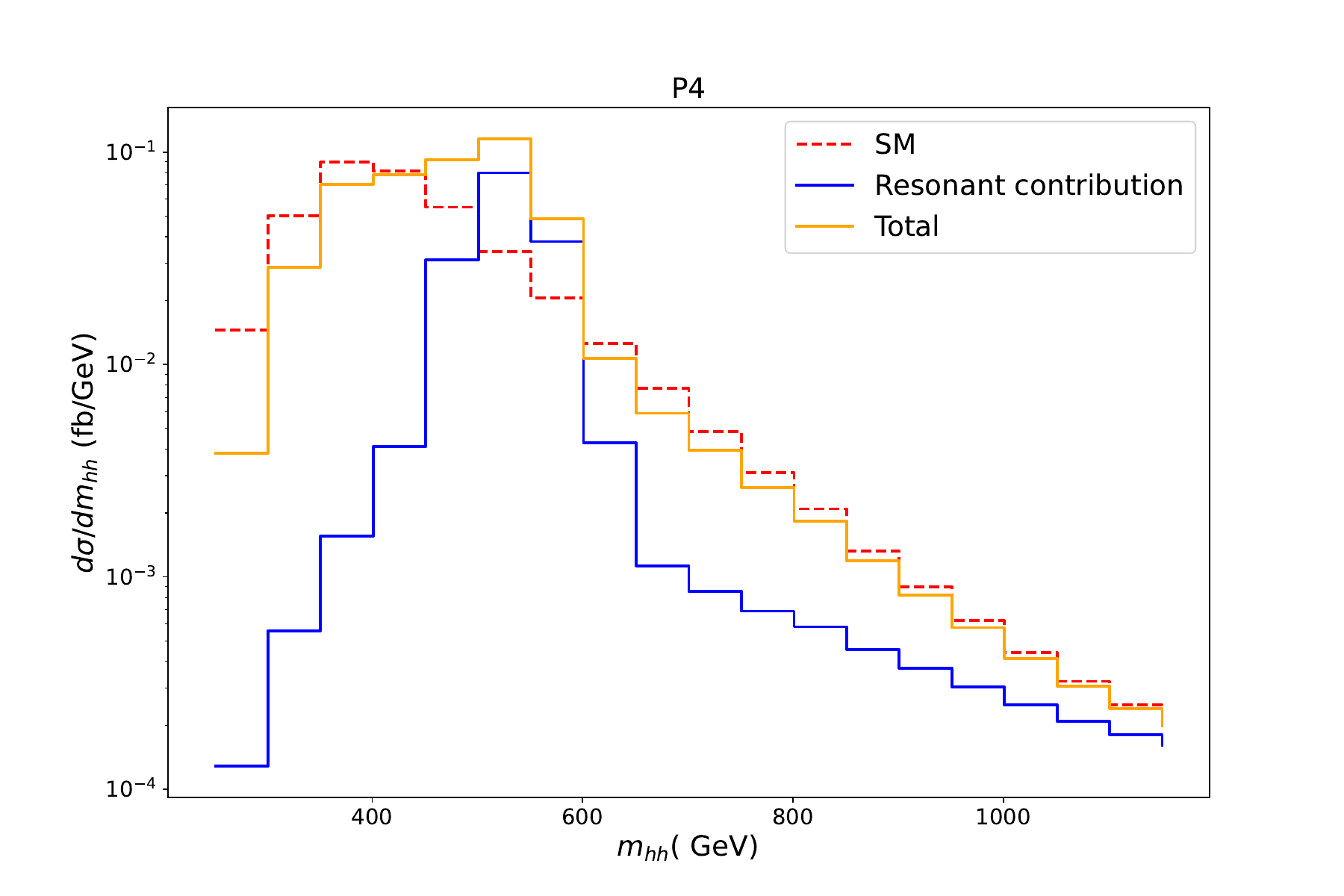}
    \includegraphics[width=0.6\linewidth]{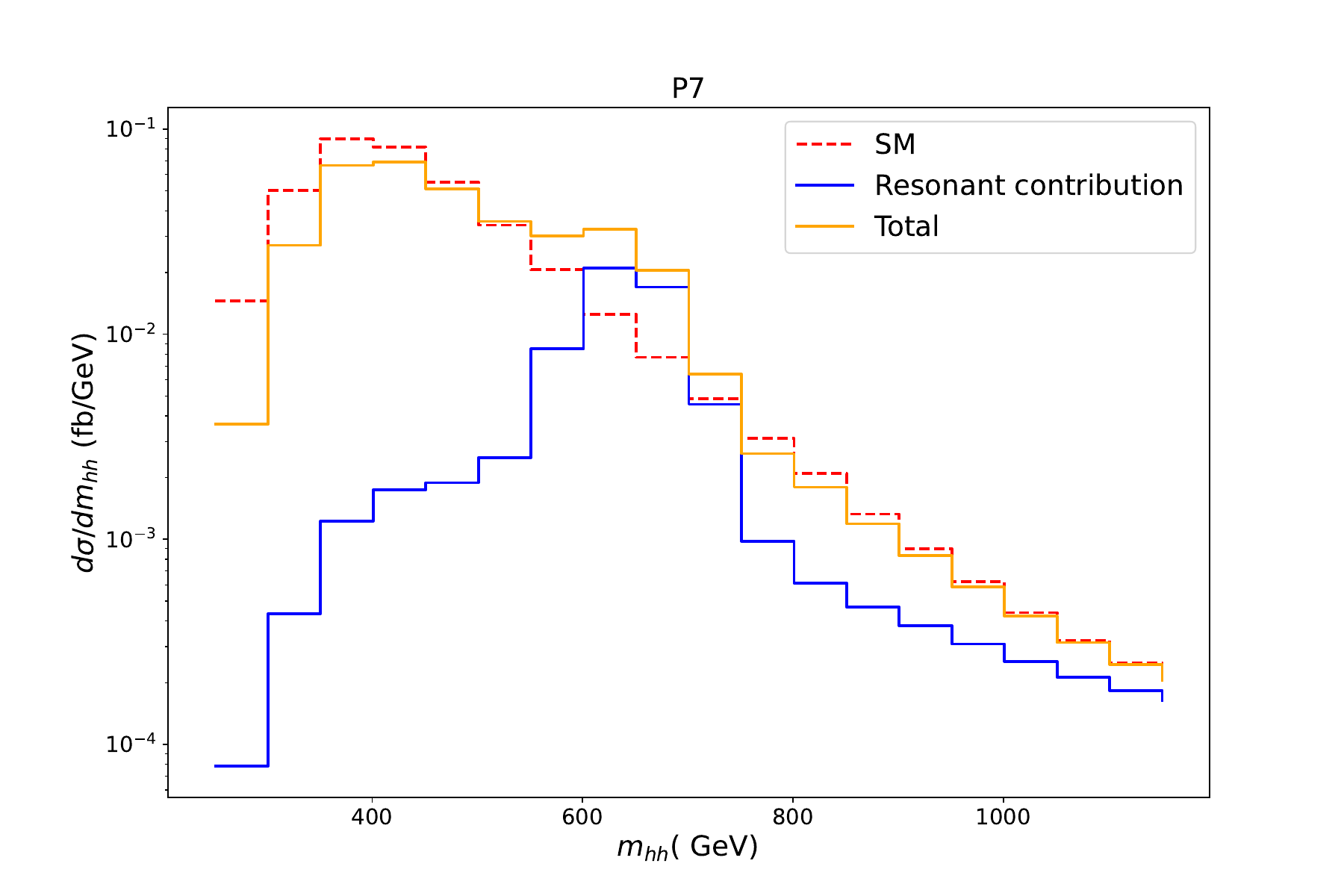}
    \caption{Differential cross section of the process $pp\to hh$ at the HL-LHC as a function of \mhh\ for the SM 
    (red dashed line), 
    compared to the distributions in P1 (region~1, upper plot), P4 (region~2, middle plot) and P7 (region~3, lower plot),
    see \protect\refta{points2}. Orange (blue) lines
    show the result of the full calculation (taking into account only the resonant diagram).
    }
    \label{fig:full-vs-res}
\end{figure}

In \reffi{fig:full-vs-res} we show the differential cross section of the process $pp\to hh$ at the HL-LHC as a 
function of \mhh\ for the SM (red dashed line), compared to the distributions in P1 (region~1, upper plot), 
P4 (region~2, middle plot) and P7 (region~3, lower plot), 
see \refta{points2}. Orange (blue) lines show the result of the full calculation (taking into account only the 
resonant diagram). For all three depicted benchmark points the pure resonant \mhh\ distribution exhibits, 
as expected, a clear peak structure around the respective $\mhh = \MH$ value. The correct \mhh\ distributions, i.e.\ 
taking correctly into account the resonant contribution, the 
non-resonant diagrams, as well as all interference contributions, have a substantially broader structure. 
For $\mhh \gsim \MH$ the \mhh\ distributions are somewhat enhanced w.r.t.\ the pure resonant result. However,
substantially larger effects of the correct full calculation are found for 
$\mhh \lsim \MH$. In P1 (region~1) the resonant peak is somewhat broadened to smaller \mhh\ values. 
For P4 (region~2) the peak is broadened 
already over several bins towards smaller \mhh\ values, where the effect becomes most pronounced for P7 (region~3). 
As argued in \citere{Heinemeyer:2024hxa}, it is plausible to conclude that such (realisticly) broadened peak structures,
deviating strongly from the
clear peak structure of the pure resonant contribution, could not be identified by the current design of the 
experimental searches. Conversely, applying a pure resonant \mhh\ distribution to the experimental analysis could 
lead to an erroneous exclusion of a parameter 
point, which in reality produces a substantially broadened \mhh\ ``peak structure''.



\section{ILC1000 Results}
\label{sec:ilc}

\begin{figure}[htb]
    \centering
    \includegraphics[scale=0.85]{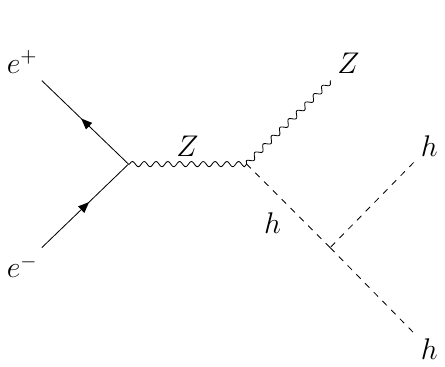}
    \includegraphics[scale=0.85]{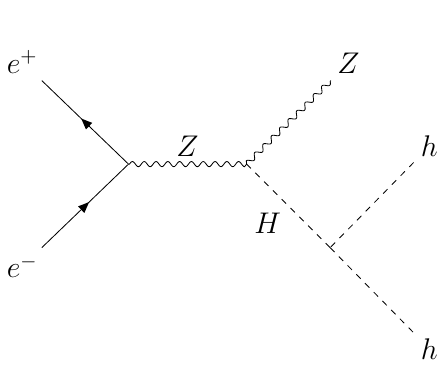} 
    \includegraphics[scale=0.85]{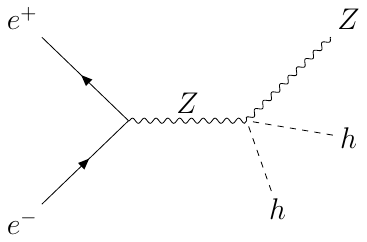}
    \hspace{7mm}
    \includegraphics[scale=0.85]{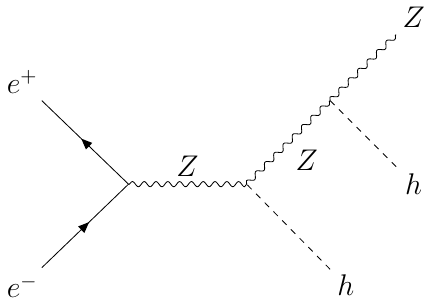}
    \caption{Generic Feynman diagrams contributing to the double Higgs-strahlung process \eeZhh\ in the RxSM. }
    \label{fig:diagrams-ee}
\end{figure}

In this section we present our results for the di-Higgs production at future high-energy $e^+e^-$ colliders.
We consider the double Higgs-strahlung channel, i.e.\ \eeZhh, which is the dominant production channel of two SM-like
Higgs bosons up to a center-of-mass energy slightly above 1~TeV. 
The Feynman diagrams that contribute to this process at tree level are shown in \reffi{fig:diagrams-ee}.
In particular, our study focuses on the effects induced by the two upper diagrams, since these are the ones containing the triple Higgs couplings $\lahhh$ (upper left) and $\lahhH$ (upper right diagram).

Similarly to the previous HL-LHC study, we employ the differential cross section distributions of the invariant mass 
of the final state Higgs-boson pair, \mhh, to study the effects of the two THCs involved in the cross section prediction.
The contributions proportional to $\lahhh$ enter via a non-resonant diagram, similar to the SM case.
Correspondingly, the largest effects of \lahhh\ are expected at low values of \mhh, close to the kinematic threshold.
On the other hand, the contributions proportional to the BSM THC, \lahhH, enter via a (potentially) resonant diagram 
mediated by the heavy Higgs boson $H$.
Therefore, the sensitivity to \lahhH\ could be accessed by detecting a resonance structure in the invariant mass 
distribution around $\mhh = \MH$.


\subsection{Calculation of \boldmath{$e^+e^- \to Zhh$}}
\label{sec:eeZhh}

We compute the unpolarized cross section for the double Higgs-strahlung process at the tree-level with the help 
of the public code~{\tt Madgraph5\textunderscore aMC v3.5.7}~\cite{Alwall:2014hca}. 
The input model file of the RxSM required by {\tt Madgraph} was obtained with the {\tt Mathematica} package 
{\tt SARAH-4.15}~\cite{Staub:2013tta}.
We compute the cross section for the ILC operating at a center-of-mass energy of $\sqrt{s} = 1000 \gev$ 
(ILC1000)~\cite{ILC:2013jhg,Moortgat-Pick:2015lbx,Bambade:2019fyw}.
The choice for the large center-of-mass energy is due to the large values of $\MH$ in our selected benchmark points, 
see \refta{points2}. 
In this work we assume an integrated luminosity of 8~ab$^{-1}$, as projected for the ILC1000~\cite{Bambade:2019fyw} 
(neglecting the possibility of polarized beams).

Using our computational setup, we obtain a prediction for $\sig(\eeZhh)$ of 0.12~fb in the SM for a 
center-of-mass energy of 1~TeV. 
At $\sqrt{s} = 500 \gev$ we find $\sig(e^+e^- \to Zhh) = 0.16 \fb$. For this center-of-mass energy
it is expected to have a discovery of the di-Higgs-strahlung process at the $8\sig$ level for an integrated
luminosity of $4 \iab$ (combining several polarization runs), corresponding to an experimental 
uncertainty of 16.8\%~\cite{Durig:2016jrs}.
Applying a simple scaling with the number of events, this yields a relative uncertainty of the cross section at 
$\sqrt{s} = 1000 \gev$ of $\sim 10\%$ for an integrated luminosity of $8\iab$,
which corresponds to a significance of close to $13\sigma$.

\begin{figure}[htb!]
\centering
\includegraphics[scale=0.8]{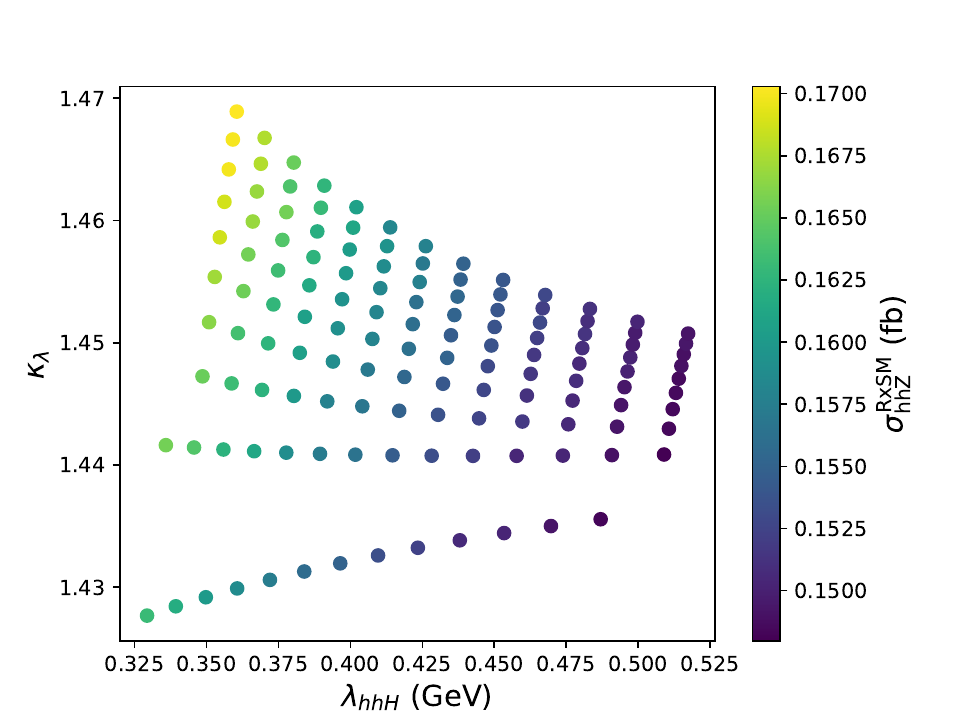}
\captionsetup{font=small} 
\caption{$\kala$-$\lahhH$ plane as a function of the total cross section for the process $e^{+}$$e^{-}$ $\to$ $Zhh$ 
at the ILC1000 in our benchmark plane.}
\label{fig:kala1000}
\end{figure}

In \reffi{fig:kala1000} we show the prediction of $\sig(\eeZhh)$ in the RxSM in the \lahhH-\kala\ benchmark plane, 
see \refse{sec:planes}. It ranges from $\sim 0.148 \fb$ for smaller \kala\ and larger \lahhH, up to $\sim 0.173 \fb$ 
for larger \kala\ and smaller \lahhH. 
Rescaling the expected precision with the enhanced cross section in the RxSM w.r.t.\ the SM prediction, this yields a 
difference of the RxSM prediction in our benchmark plane between $\sim 2\,\sig$ up to $\sim 4\,\sig$ w.r.t.\ the SM 
applying \refeq{significance}, but in the case of an $e^+e^-$ collider). 
Another observation can be made for the contribution of the heavy Higgs-boson resonance. The size of the $H$-resonance 
contribution becomes smaller with increasing \lahhH. The reason behind is the fact that in our benchmark plane 
the product of $(\sin\theta \cdot \lahhH)$ is approximately constant.
Larger \lahhH, however, corresponds to larger $\MH$, leading to 
an overall suppression of $\sig(e^+e^- \to ZH) \times \br(H \to hh)$ for larger \lahhH, as can be seen in 
the blue curves in \reffis{fig:P3P4-ee-mhh} and \ref{fig:P7P8-ee-mhh} below. On the other hand, the cross section 
leaving out the $H$-resonance contribution remains nearly constant over the plane, as can be observed in the green 
curves in \reffis{fig:P3P4-ee-mhh} and 
\ref{fig:P7P8-ee-mhh}, below. In combination with the corresponding interference effects, the overall contribution of the 
heavy Higgs-boson resonance to the total cross section decreases with increasing \lahhH, leading to smaller 
$\sig(\eeZhh)$ for larger \lahhH\ as observed in \reffi{fig:kala1000}.
A more detailed analysis of the THC dependencies will be given in the next subsection.
But it is interesting to note that a parameter space of the RxSM that was identified to yield a strong FOEWPT and 
is favorable for the di-Higgs production at the LHC~\cite{paper} yields possibly detectable deviations from the SM 
also at $e^+e^-$ colliders. 

\bigskip
While we also comment on the effects of \lahhh\ on the total and the differential cross sections in the next subsection, 
a major focus of this work is to study the potential sensitivity of the ILC1000 to the THC, \lahhH,  via an analysis of
the $H$ resonance structure. In order to take into account in more detail the experimental analyses (i.e.\ detector
effects, cuts, etc.)
we focus on the main light Higgs-boson decay channel, $h \to b\bar b$, which in the SM has a BR of $\sim 0.58$. 
Consequently, the main experimental signature is given by four $b$-quark jets together with a $Z$ boson.
Therefore, following a similar strategy as in \citere{Arco:2021bvf}, we estimate the expected number of events with 
four~$b$ jets and one $Z$ boson, denoted by $\bar N$, that could be detected at the ILC1000 with the following expression:
\begin{equation}
    \bar N = N \times \left(\br(h\to b\bar b)\right)^2 \times {\cal A} \times \epsilon_b^4\,,
    \label{eq:Nbar}
\end{equation}
where $N$ is the inclusive number of $Zhh$ events predicted by the RxSM, as calculated above.
We assume a conservative $b$-tagging efficiency of $\epsilon_b=80\%$ for each final $b$-jet.
We also assume 
the SM prediction for the $h\to b\bar b$ branching ratio (in good agreement with the LHC measurements).
${\cal A}$ is our estimation of the detector acceptance after applying the following pre-selection cuts to detect the
final $4b+Z$ events (see \citere{Arco:2021bvf} for details):
\begin{equation}
p_{T}^{Z} > 20\ \text{GeV}, \quad p_{T}^{b} > 20\  \text{GeV}, \quad |\eta_b| < 2, \quad \Delta R_{bb} > 0.4\,,
\label{eq:cuts-ee}
\end{equation}
where $p_T^Z$ and $p_T^b$ are the transverse momenta of the $Z$ boson and each of the $b$ quarks, respectively, 
$\eta_b$ is the  pseudo-rapidity of each of the $b$ quarks, and $\Delta R_{bb}$ is the angular separation between 
two $b$~quarks defined by  
$\Delta R_{ij} = \sqrt{\left(\eta_i-\eta_j\right)^2+\left(\phi_i-\phi_j\right)^2}$, with $\phi$ being the azimuthal angle.
We compute the acceptance ${\cal A}$ by simulating the process \eeZhh\ with the subsequent decay $h\to b\bar b$ with and 
without the cuts defined in \refeq{eq:cuts-ee} in {\tt MadGraph} at the parton level.
Therefore, the acceptance is given by the ratio of the predicted $4b+Z$ events with and without cuts.
The obtained values of the acceptances ${\cal A}$ for the studied BPs are given in \refta{tab:Acc}. 

\begin{table}[h]
\centering
\begin{tabular}{c|cccccccc}
 & P1 & P2 & P3 & P4 & P5 & P6 & P7 & P8 \\ \hline
\textbf{$\cal A$} & 0.7308 & 0.7382 & 0.7376 & 0.7638 & 0.7725 & 0.7625 & 0.7963 & 0.7982
\end{tabular}
\caption{Detector acceptances at the ILC1000 (see text) for the benchmark points defined in \refta{points2}.}
    \label{tab:Acc}
\end{table}

To evaluate the potential sensitivity to the $H$ resonant peak, and therefore to $\lahhH$, of the ILC1000 we again 
use the ``theoretical estimator'' $R$ defined in \refeq{rpar}.
Analogously to the HL-LHC case, $\bar{N}_{i}^{R}$ and $\bar{N}_{i}^{C}$
denote the expected events, as defined in \refeq{eq:Nbar},
in the $i$th bin from the purely resonant diagrams (the one mediated by $H$ and proportional to $\lahhH$) and 
the non-resonant ones, respectively.%
\footnote{Note that the definition of $\bar N^C$ slightly differs from the one used in the HL-LHC analysis, where
the SM curve was employed for $N^C$. However, this has a minor numerical impact.}%
~In contrast to the HL-LHC, in the case of the ILC1000 we define the signal region such that the difference between the 
resonant and non-resonant number of expected events is at least two, i.e.,
\begin{equation}
|\bar{N}_i^{R} - \bar{N}_i^{C}| > 2.
\label{eq:minnumev}
\end{equation}

Here it should be kept in mind that, similar to the HL-LHC case, the ``theoretical estimator'' $R$ gives an idea of 
how prominent the $H$ resonance peak is relative to the continuum contributions from the non-resonant diagrams. 
Large values of $R$ indicate more accessible $H$ resonance peaks at the ILC1000, which implies higher chances of 
obtaining potential experimental information about the $\lahhH$ coupling. As in the case of the HL-LHC, $R$ does 
not correspond to a true ``experimental significance''.


\subsection{Analysis of \boldmath{\mhh}}
\label{sec:mhh-ee}

\begin{figure}[htb]
\begin{subfigure}{0.48\textwidth}
    \includegraphics[scale=0.35]{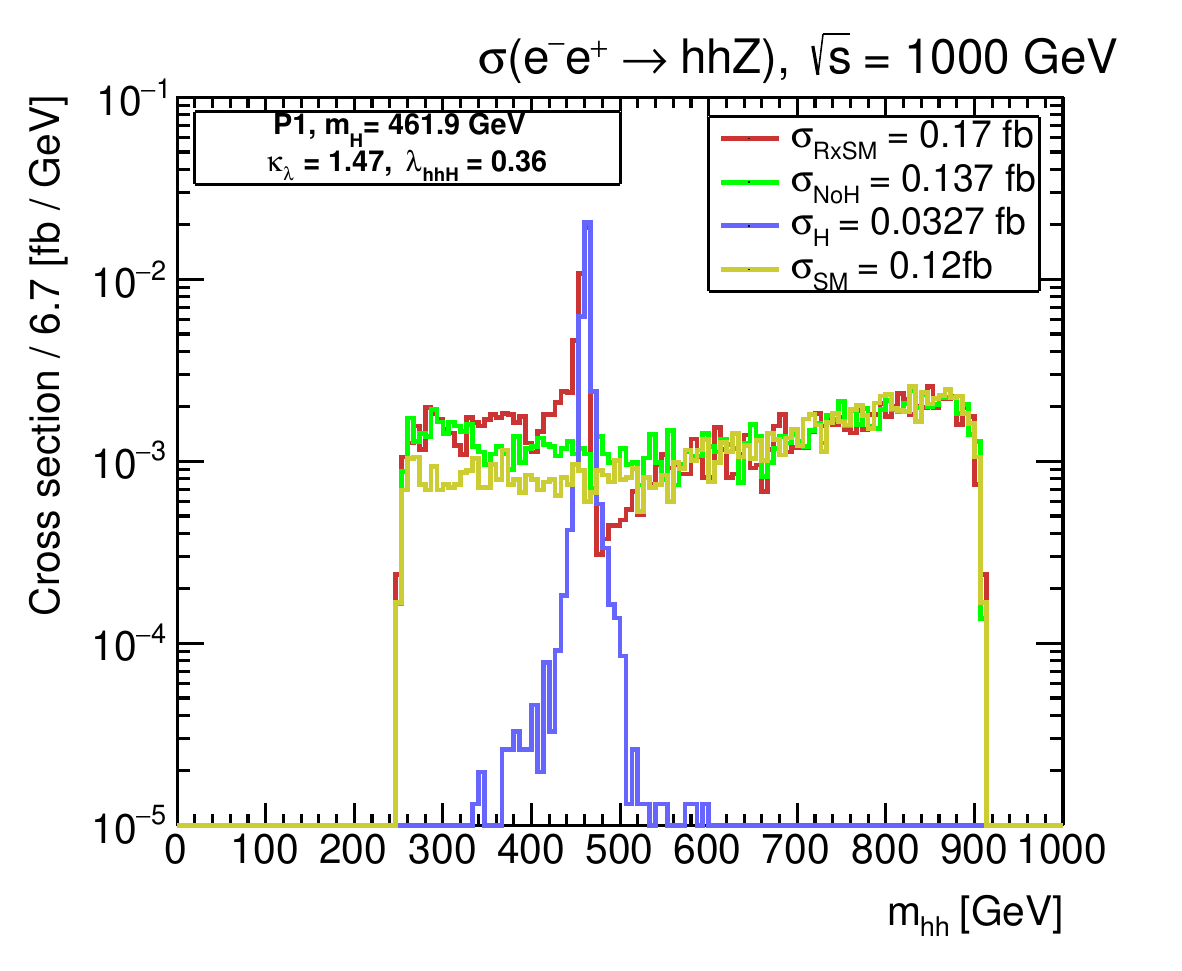}
\end{subfigure}
\hfill
\begin{subfigure}{0.48\textwidth}
    \includegraphics[scale=0.35]{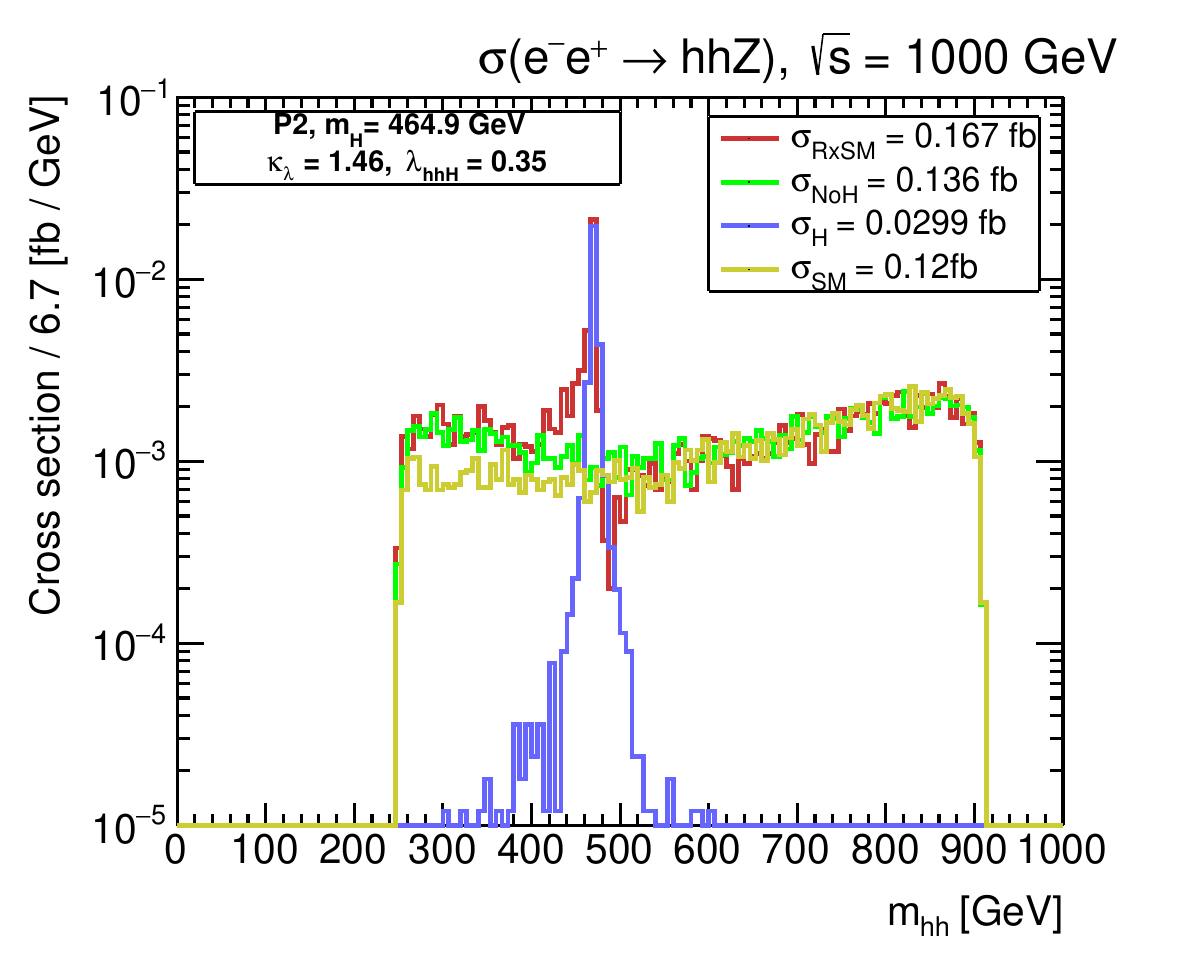}
\end{subfigure}
\begin{subfigure}[b]{0.48\textwidth}
\includegraphics[scale=0.35]{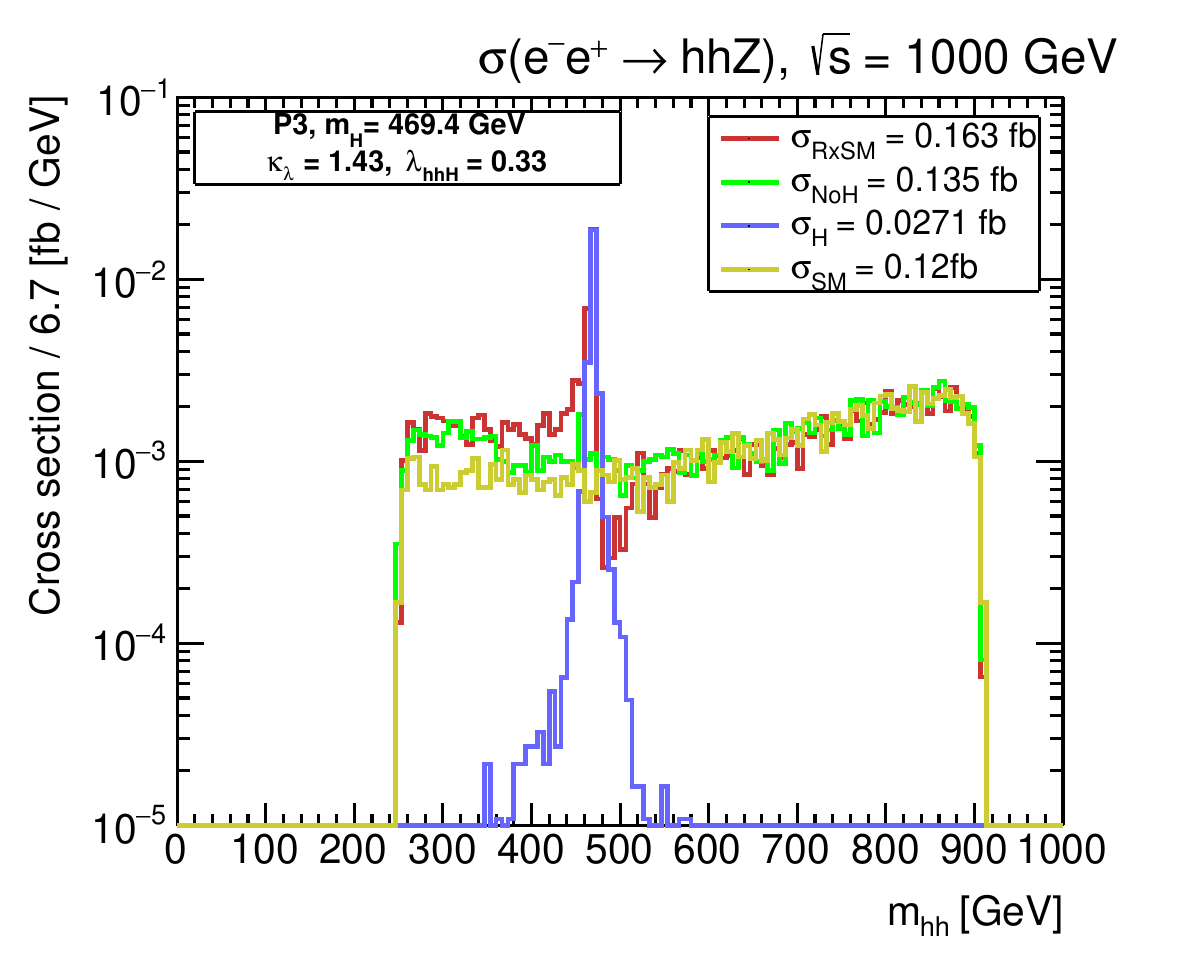}
  \end{subfigure}
  \hfill
  \begin{subfigure}[b]{0.48\textwidth}
\includegraphics[scale=0.35]{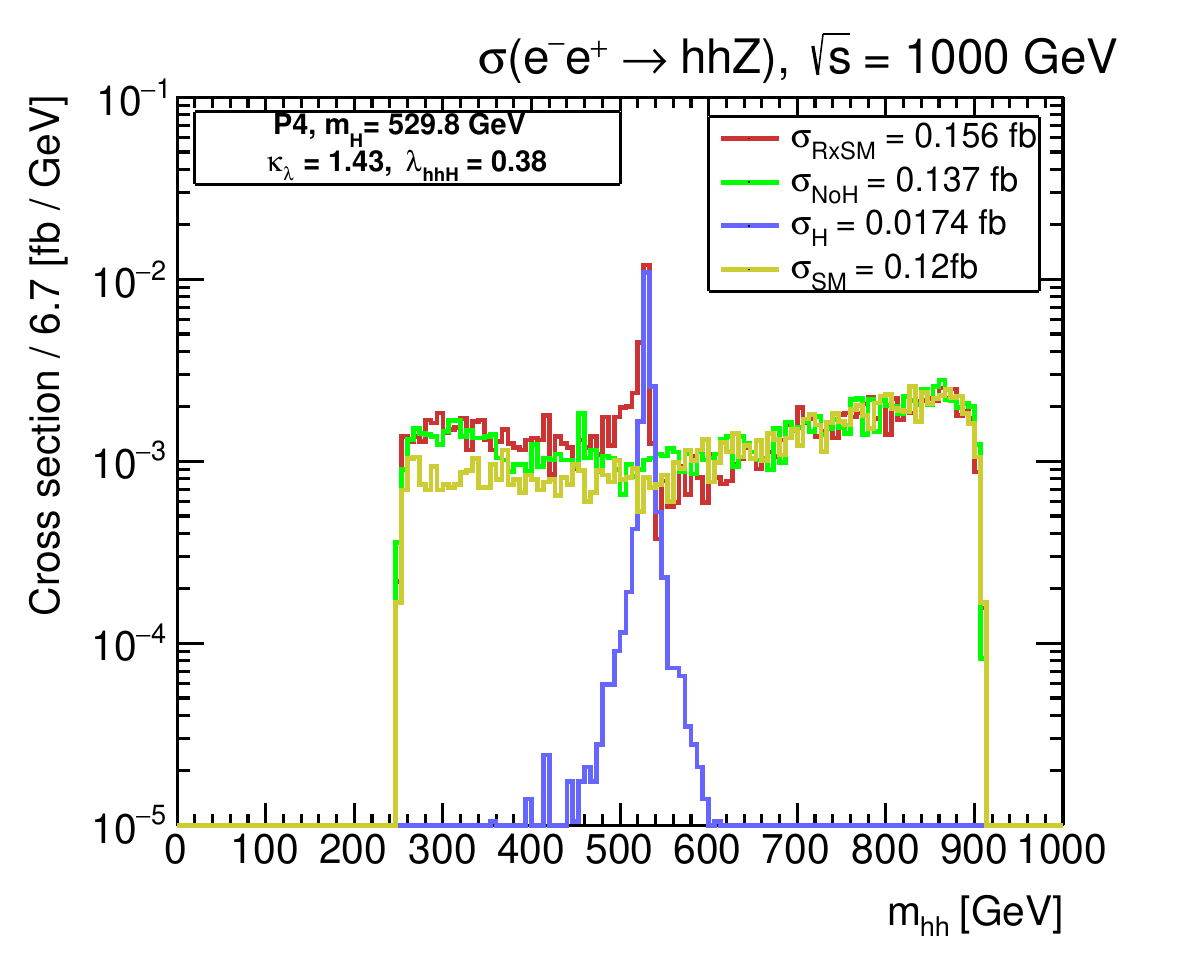}
  \end{subfigure}
  \captionsetup{font=small}
  \caption{Differential cross section of the process $e^{+}$$e^{-}$ $\to$ $Zhh$ at the ILC1000 as a 
  function of $m_{hh}$ for the 
  SM, for the pure heavy Higgs resonant contribution (blue), for the non-resonant contributions (green), and for 
  the full RxSM calculation (red) for benchmark points P1, P2, P3,  and P4.
  }
\label{fig:P3P4-ee-mhh}
\end{figure}

\begin{figure}[htb]
\begin{subfigure}{0.48\textwidth}
    \includegraphics[scale=0.35]{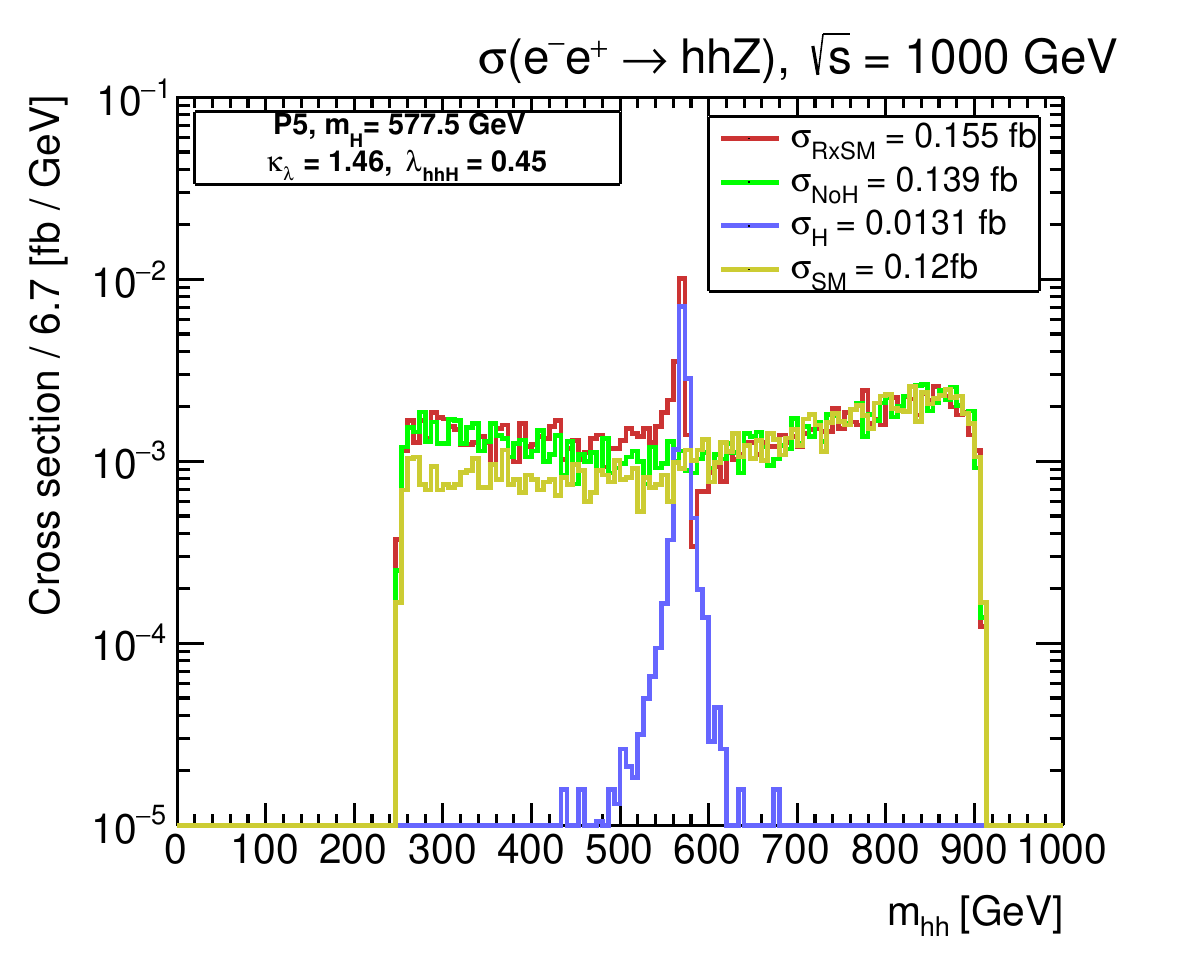}
\end{subfigure}
\hfill
\begin{subfigure}{0.48\textwidth}
    \includegraphics[scale=0.35]{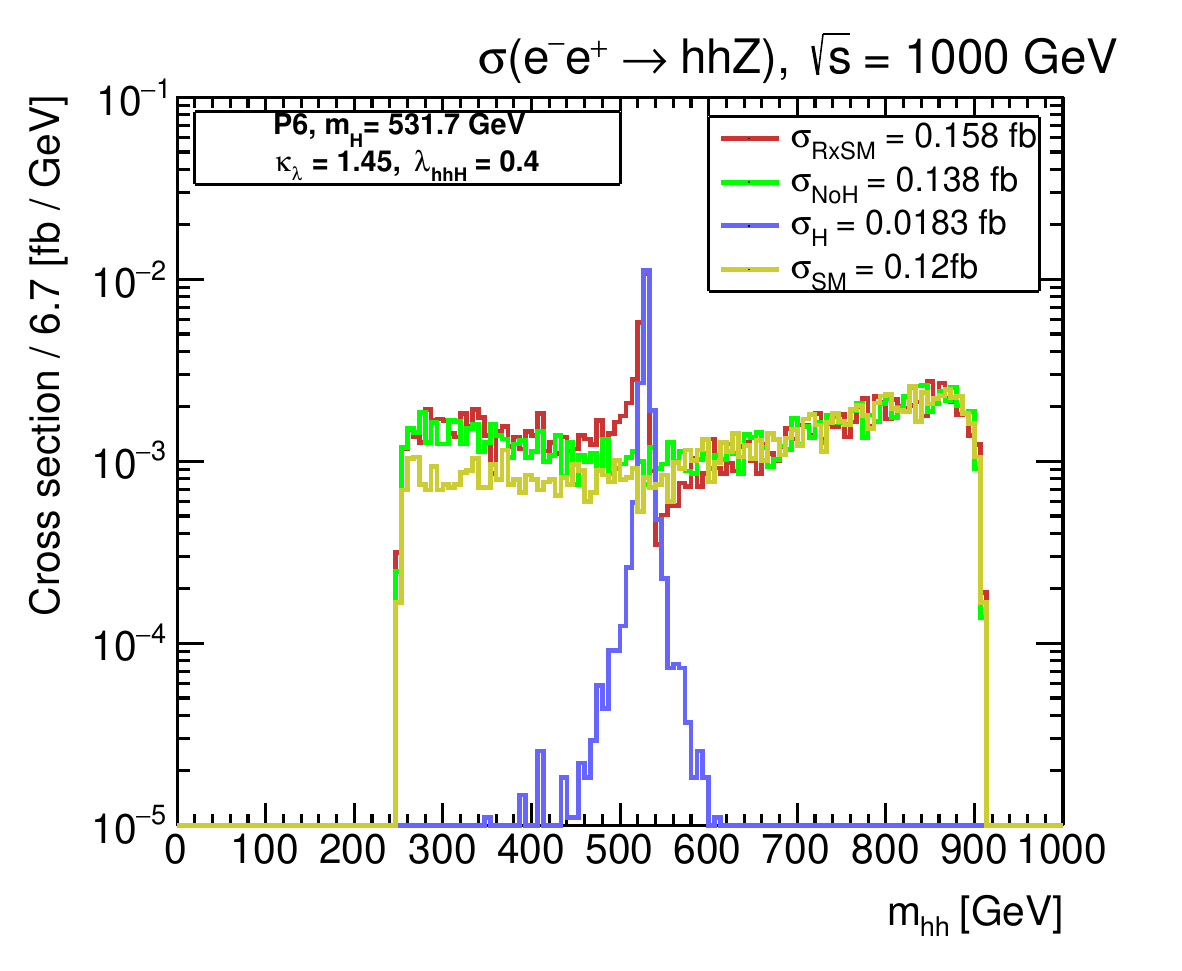}
\end{subfigure}
\begin{subfigure}[b]{0.48\textwidth}
\includegraphics[scale=0.35]{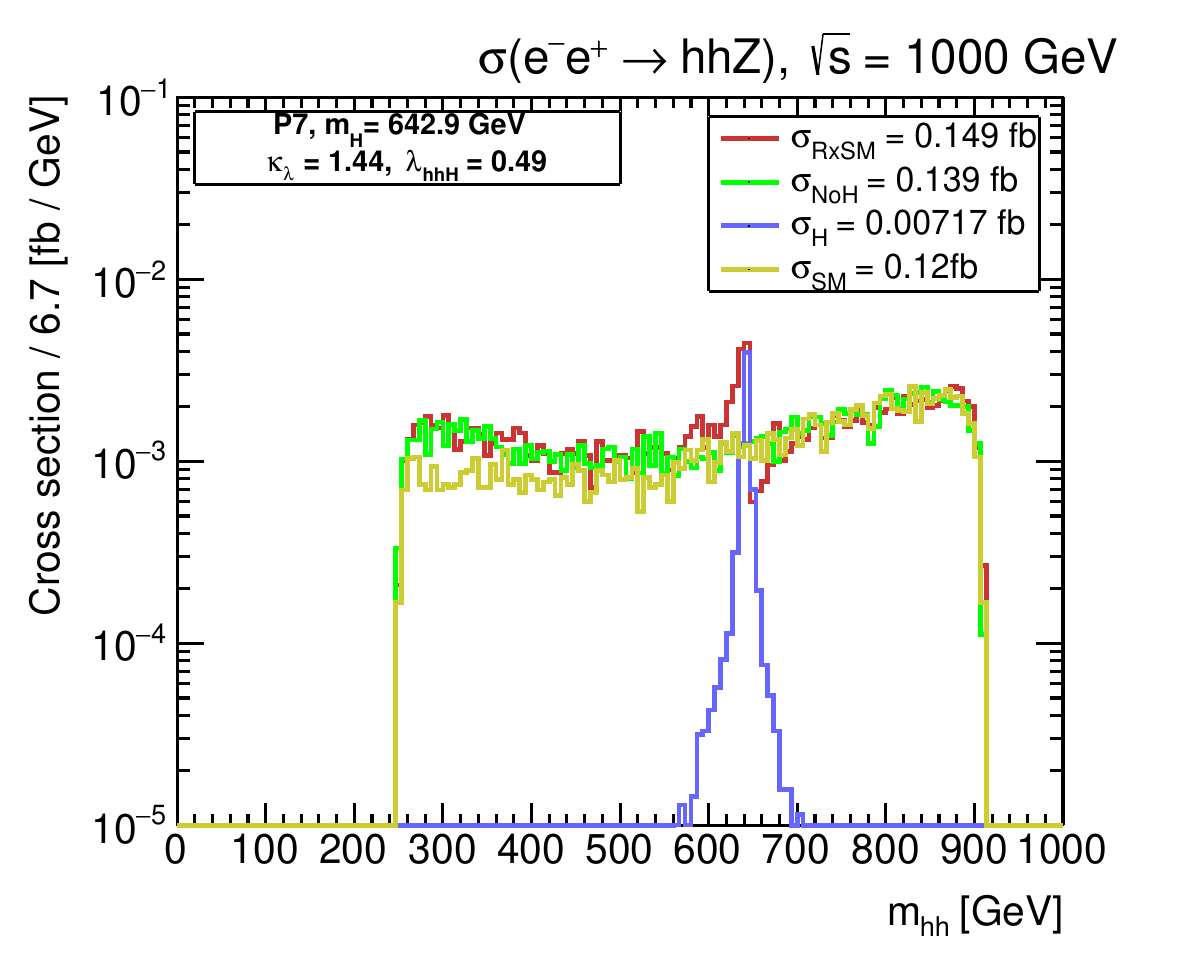}
  \end{subfigure}
  \hfill
  \begin{subfigure}[b]{0.48\textwidth}
\includegraphics[scale=0.35]{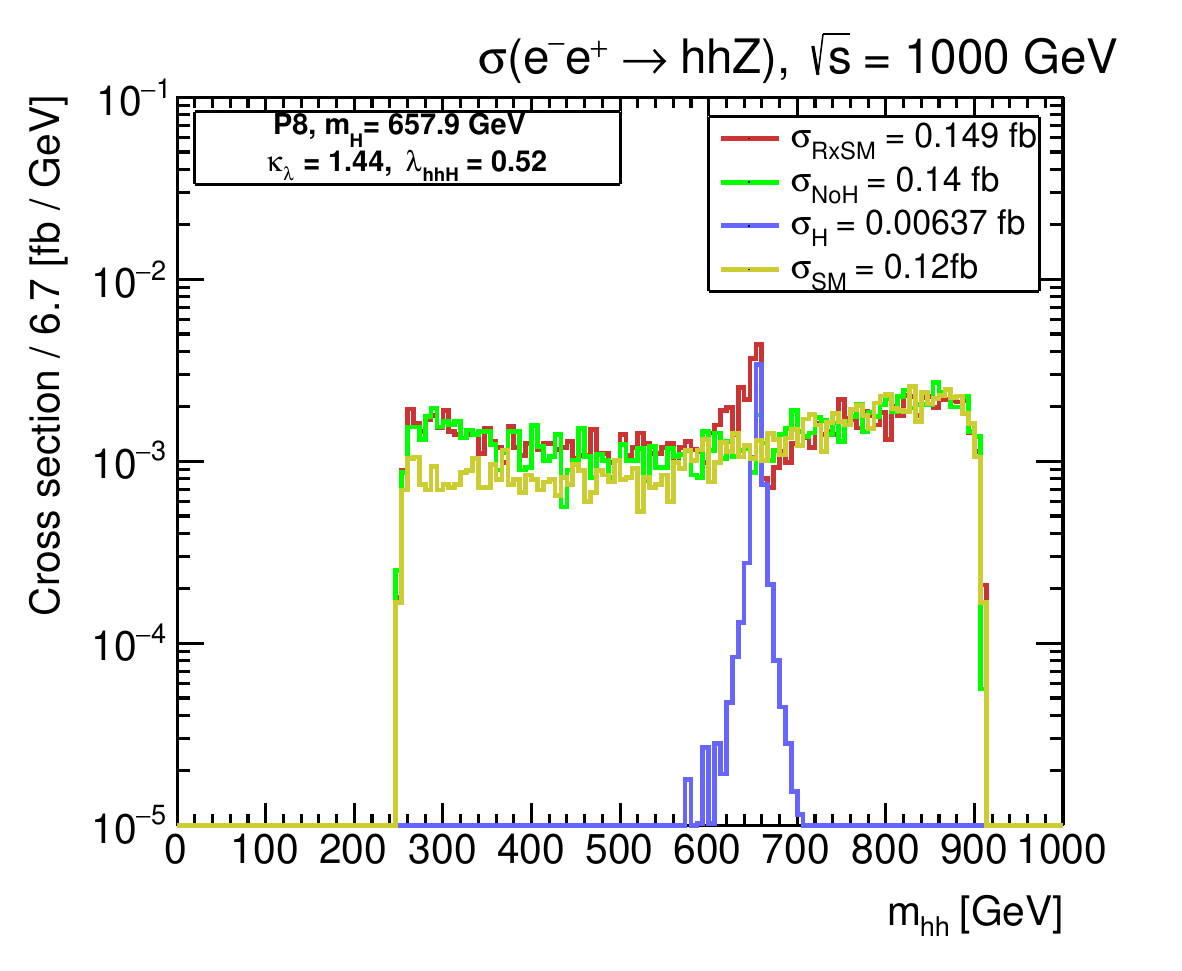}
  \end{subfigure}
  \captionsetup{font=small}
  \caption{\mhh\ distribution for benchmark points P5 - P8, with the color coding as in \protect\reffi{fig:P3P4-ee-mhh}.}
\label{fig:P7P8-ee-mhh}
\end{figure}

In this subsection we analyze the differential \mhh\ distributions for the eight benchmark points as defined in 
\refta{points2} at the ILC1000. In \reffi{fig:P3P4-ee-mhh} and \reffi{fig:P7P8-ee-mhh}, we present the results 
for P1--P4 and P5--P8, respectively.
In each plot the red curve corresponds to the full RxSM prediction (\sigeeRx), 
whereas the green and and the blue curves show, respectively, the results leaving out the resonant $H$ diagram (\sigeeNoH)
and taking into account the resonant $H$ diagram only (\sigeeH). For comparison, the green curve indicates
the SM result (\sigeeSM). 
The values for the total cross sections are given in the legends, as well as the values of $\MH$, \kala\ and \lahhH\ 
for each benchmark point.
The binning of $6.7 \gev$ is chosen according to \refeq{eq:minnumev}.

Each benchmark point yields the same qualitative features, but differs in the location of the resonance structure,
determined by the respective $\MH$ value, the ``height'' of the resonance, determined by 
$\sin\theta \cdot \lambda_{hhH}$ and $\MH$, where the same holds for \sigeeH.
For the two RxSM cross sections involving the $h$ exchange (\sigeeRx, \sigeeNoH) \kala\ plays a significant role.
The effects of $\kala \sim 1.45$ are best visible in the comparison of \sigeeSM\ and \sigeeNoH. 
As expected (and also observed in \citere{Arco:2021bvf}), the enhancement \wrt the SM is most pronounced for small \mhh\ 
and leads to an enhancement of the differential cross section prediction (corresponding to an enhanced total cross section,
as known for $\kala > 1$ in the \eeZhh\ channel). This effect is nearly identical for all eight benchmark points, as the 
value of $\kala$ varies only slightly over the whole benchmark plane.

The differential cross section that would be given by the pure resonance contribution is shown by the blue curves 
in \reffis{fig:P3P4-ee-mhh} and (\ref{fig:P7P8-ee-mhh}). One can observe that this cross section is largest for 
smaller \lahhH. As discussed above, this is 
due to the fact that in our benchmark plane larger \lahhH\ corresponds to larger $\MH$, while $(\sin\theta \cdot \lahhH)$ 
remains approximately constant, leading to 
an overall suppression of the heavy Higgs-boson resonant contribution for larger \lahhH. 
Taking into account the interference with the
non-resonant diagrams, this yields a peak-dip structure around $\mhh = \MH$ as clearly visible in the red curves. 
However, corresponding to the size of the $H$-resonance contribution, also the peak-dip structure is strongest for 
smaller \lahhH, which can be seen best in the $R$ values:
for each of the points we have evaluated the $R$ value according to \refeq{rpar}, which are summarized in
\refta{tab:sensitivity}. 
For an easier comparison, we also repeat in that table the values of $\MH$ and \lahhH\ for each benchmark point.
It can be observed that within our benchmark plane (which is favored by the phenomenology of a strong FOEWPT) 
smaller values of \lahhH\ lead (to a good approximation)
to a stronger ``signal'' of the resonant $H$ contribution and thus to a larger $R$ value -- 
in agreement with our discussion above. 
The overall substantially lower values of $R$ as 
compared to the HL-LHC result on the one hand from a more realistic set-up including cuts etc., 
and on the other hand from the overall lower number of events at the $e^+e^-$ collider. 

\begin{table}[htb!]
\centering
\begin{tabular}{c|cccccccc}
 & P1 & P2 & P3 & P4 & P5 & P6 & P7 & P8 \\ \hline
\textbf{$R$} & 13.59 & 14.35 & 10.54 & 7.16 & 5.75 & 7.26 & 3.39 & 3.70 \\
$\MH$ [GeV] & 461.9 & 470.8 & 469.4 & 530.9 & 575.1 & 529.6 & 642.5 & 656.1 \\
\lahhH & 0.36 & 0.35 & 0.33 & 0.38 & 0.45 & 0.40 & 0.49 & 0.52
\end{tabular}
\caption{Values of the sensitivity $R$ for the eight benchmark points, see \protect\refta{points2}.}
\label{tab:sensitivity}
\end{table}


\section{Conclusions}
\label{sec:conclusions}

In this work we have analyzed the impact of triple Higgs couplings on the production cross section
of two $\sim 125 \gev$ Higgs bosons at the HL-LHC and the ILC1000. 
We have chosen the Higgs singlet extension of the SM without $Z_2$ symmetry, the RxSM, as an example framework.
We have focused on a benchmark plane that is phenomenologically 
favored, as it yields a strong FOEWPT 
in the early universe (based on the original analysis of \citere{paper}), a key ingredient 
to fulfill one of the three Sakharov conditions required for EW baryogenesis to explain the baryon asymmetry 
of the universe. We have ensured that the plane under
consideration is in agreement with all theoretical and experimental constraints. By the requirement of the FOEWPT
the benchmark plane is not in the alignment limit. The main idea of our work is to analyze the effect of the THC, 
encoded in $\kala \equiv \lahhh/\laSM$, which is found to be $\kala \sim 1.45$, i.e.\ far away from the SM value. 
The second focus is on the impact of the BSM THC \lahhH, which enters via a heavy Higgs-boson exchange with the subsequent 
decay $H \to hh$, in the di-Higgs production cross section both at the HL-LHC and at the ILC.

For the HL-LHC we calculated $\sig(gg \to hh)$ as well as ${\rm d}\sig(gg \to hh)/{\rm d}\mhh$ with the help of the
code \texttt{HPAIR}, adapted to the RxSM. Within our phenomenologically favored benchmark plane the total cross section 
can deviate by more than $3\sig$ from the SM cross section. In other parts of the benchmark plane the total cross section
would remain experimentally indistinguishable from the SM prediction. Since we are away from the alignment limit, 
this equality between the RxSM and the SM results
is due to (accidental) cancellations of the two BSM effects stemming from $\kala > 1$ and from the contribution 
of the heavy Higgs-boson resonance, inducing a dependence on \lahhH. 

This effect becomes better visible in the second part of the HL-LHC analysis, focusing on the \mhh\ distributions.
These have been evaluated for eight benchmark points, distributed over our benchmark plane. For those we take into account
a 15\% detector smearing and a $50 \gev$ binning in \mhh. The eight benchmark points are compared with the SM expectation.
A simple ``theoretical significance'', $R$, is employed (as defined and used in an 2HDM HL-LHC analysis in 
\citere{Arco:2022lai}) that allows us to estimate the ``visibility'' of the $H$-resonance peak w.r.t.\ the 
SM expectation. While this does not constitute a realistic experimental significance, 
this estimator allows to compare different benchmark points with each other (as well as collider energy and 
luminosity options). The large values found for $R$ in our eight 
benchmark points, spanning the whole plane favored by the FOEWPT, give rise to the hope that also in a realistic
experimental analysis a clear sign of the $H$-resonance peak can be observed, giving access to the BSM THC \lahhH. 

In the final step of our HL-LHC analysis in the RxSM we compare the \mhh\ distributions evaluated solely from the
resonance diagram, but neglecting the two continuum diagrams with the results from the full calculation 
(i.e.\ taking into account all contributing diagrams and in particular the corresponding interferences). 
This approach of neglecting the continuum diagrams is currently taken by the experimental
collaborations, ATLAS and CMS, to obtain their results for resonant di-Higgs production. In the context of the 2HDM, 
in \citere{Heinemeyer:2024hxa} it was demonstrated that this approach can lead to the erroneous exclusion of 
parameter points. In our RxSM analysis, comparing the \mhh\ distributions either neglecting or including the 
continuum diagrams, we find (as expected) strong and relevant differences between the two types of distributions. 
In particular, in the full calculation the ``resonance peak'' is substantially  broadened. This sheds severe doubts 
that the current experimental data from resonant di-Higgs production at the LHC can readily be applied to  our 
phenomenologically favored benchmark plane in the RxSM.

\smallskip
The analysis of the RxSM benchmark plane featuring a strong FOEWPT was subsequently extended 
to future $e^+e^-$ colliders. As a particular
example we focused on the ILC with a center-of-mass energy of $\sqrt{s} = 1000 \gev$, the ILC1000. 
As an integrated luminosity we assume
$8 \iab$. The center-of-mass energy is required since our benchmark points have $\MH$ values that yield a 
resonant contribution (and thus possibly access to \lahhH) only for $\sqrt{s} \gsim 600 \gev$ (depending on $\MH$), 
making the ILC1000 the preferred option.
From the total ILC1000 cross section alone, in our benchmark plane we find a deviation w.r.t.\ the SM prediction between 
$\sim 2$ and $4\,\sig$. Also for the ILC1000 we calculated the \mhh\ distributions for the eight benchmark points,
evaluating the ``theory estimator'' $R$ adapted to the $e^+e^-$ case (following a corresponding 2HDM analysis in
\citere{Arco:2022xum}). In this case also experimental cuts were included to take into account detector efficiencies 
etc.\ (again following \citere{Arco:2022xum}). 
We find that for the eight benchmark points, to a good approximation, smaller values of \lahhH\ lead 
to a stronger ``signal'' of the resonant $H$-exchange contribution and thus to a larger $R$ value. 
Overall substantially lower values of $R$ as compared to the HL-LHC result are found. 
This results on the one hand from a more realistic set-up including cuts etc., and on the other hand from the 
overall lower number of events at the $e^+e^-$ collider. 
Nevertheless, as in the HL-LHC case this gives rise to the hope that also in a realistic experimental analysis a 
clear sign of the $H$-resonance peak can be observed, giving access to the BSM THC \lahhH\ at the ILC1000.

Overall, we conclude that within the RxSM, depending on the values of the underlying Lagrangian
parameters, a sizable resonant $H$ contribution to the di-Higgs
production cross section of two SM-like Higgs bosons can leave possibly visible effects in the \mhh\ distribution. 
This applies to the HL-LHC or to a future $e^+e^-$ collider (taking the ILC1000 as a concrete example).
This would pave the way for a first determination of a BSM THC, a step that is crucial for the reconstruction of the Higgs
potential of the underlying BSM model.


\subsection*{Acknoledgements}

We thank 
M.~Ramsey Musolf
for helpful discussions.
F.A. acknowledge support by the Deutsche Forschungsgemeinschaft (DFG, German Research Foundation) 
under Germany‘s Excellence Strategy –-- EXC 2121 “Quantum Universe” –-- 390833306. This work has been partially 
funded by the Deutsche Forschungsgemeinschaft (DFG, German Research Foundation) - 491245950.
S.H.\ acknowledges partial financial support by the Spanish Research Agency (Agencia Estatal de Investigaci\'on) 
through the grant IFT Centro de Excelencia Severo Ochoa No CEX2020-001007-S funded by MCIN/AEI/10.13039/501100011033. 
The work of S.H.\ was also supported by the Grant PID2022-142545NB-C21 funded by
MCIN/AEI/10.13039/501100011033/ FEDER, UE.
The work of M.M.\ has been supported by the BMBF-Project 05H24VKB.
The work of A.V.S.\ is supported is in part by the Deutsche Forschungsgemeinschaft (DFG, German Research Foundation) 
Emmy Noether Grant No. BR 6995/1-1, and under Germany‘s Excellence Strategy --- EXC 2121 ``Quantum Universe'' --–
390833306, and in part by the Deutsche Forschungsgemeinschaft (DFG, German Research Foundation) --- 491245950.


\end{document}